\documentclass[fleqn,usenatbib]{mnras}

\usepackage{newtxtext,newtxmath}

\usepackage[T1]{fontenc}
\usepackage{graphicx}
\usepackage{amsmath}
\usepackage{orcidlink}

\defcitealias{amarsi2019}{A19}
\defcitealias{kobayashi2020}{K20}

\title[ChemZz I: Oxygen-to-iron across cosmic time]{ChemZz I: Comparing Oxygen and Iron Abundance Patterns in the Milky Way, the Local Group and Cosmic Noon}

\author[S. Monty et al.]{
\noindent Stephanie Monty$^{1}$\thanks{E-mail: sm2744@cam.ac.uk} \orcidlink{0000-0002-9225-5822}, 
Allison L. Strom$^{2,3,4}$ \orcidlink{0000-0001-6369-1636}, 
Thomas M. Stanton$^{5}$ \orcidlink{0000-0002-0827-9769}, 
Martyna Chru\'sli\'nska$^{6,7}$ \orcidlink{0000-0002-8901-6994},
\newauthor
Fergus Cullen$^{5}$ \orcidlink{0000-0002-3736-476X},
Chiaki Kobayashi${^8}$\orcidlink{0000-0002-4343-0487},
Tjitske Starkenburg$^{2,3,4}$\orcidlink{0000-0003-2539-8206},
Souradeep Bhattacharya$^{8}$\orcidlink{0000-0003-4594-6943},
\newauthor
Jason L. Sanders$^{9}$\orcidlink{0000-0003-4593-6788},
Mark Gieles$^{10, 11}$\orcidlink{0000-0002-9716-1868}
\\
$^{1}$Institute of Astronomy, University of Cambridge, Madingley Road, Cambridge CB3 0HA, UK \\
$^{2}$Department of Physics and Astronomy, Northwestern University, 2145 Sheridan Road, Evanston, IL 60208, USA \\
$^{3}$Center for Interdisciplinary Exploration and Research in Astrophysics (CIERA), Northwestern University, 1800 Sherman Avenue,
Evanston, IL 60201, USA \\
$^{4}$NSF-Simons AI Institute for the Sky (SkAI), 172 E. Chestnut St., Chicago, IL 60611, USA\\
$^{5}$Institute for Astronomy, University of Edinburgh, Royal Observatory, Edinburgh, EH9 3HJ, UK \\
$^{6}$European Southern Observatory, Karl-Schwarzschild-Str. 2, 85748 Garching, Germany \\
$^{7}$Max Planck Institute for Astrophysics, Karl-Schwarzschild-Str. 1, 85748 Garching, Germany\\
$^{8}$Centre for Astrophysics Research, Department of Physics, Astronomy and Mathematics, University of Hertfordshire, Hatfield, AL10 9AB, UK\\
$^{9}$ Department of Physics and Astronomy, University College London, London WC1E 6BT, UK\\
$^{10}$ ICREA, Pg. Lluís Companys 23, E08010 Barcelona, Spain\\
$^{11}$ Institut de Ciències del Cosmos (ICCUB), Universitat de Barcelona (IEEC-UB), Martí i Franquès 1, E08028 Barcelona, Spain
}
\date{Accepted XXX. Received YYY; in original form ZZZ}

\pubyear{\the\year{}}

\begin{document}
\label{firstpage}
\pagerange{\pageref{firstpage}--\pageref{lastpage}}
\maketitle

\begin{abstract}
Our understanding of the chemical evolution of galaxies has advanced through measurements from both distant galaxies across redshift, and our own Milky Way (MW). To form a comprehensive picture, it is essential to unify these constraints, placing them on a common scale and parlance and to understand their systematic differences. In this study, we homogenize oxygen and iron measurements from star-forming galaxies at Cosmic Noon ($z{\sim}2-3$) with resolved stellar abundances from the Local Group. The MW is divided into four components, assuming the outer halo is dominated by debris from the Gaia-Sausage-Enceladus (GSE) progenitor. After converting all abundances to a common Solar scale, we identify clear $\alpha$- and iron-enhancement trends with mass in the $z{\sim}2-3$ galaxies and find good agreement between these galaxies and the MW high-$\alpha$ disc in [O/Fe] vs. [Fe/H]. We also find excellent agreement between the [O/Fe] trends seen in the MW high- and low-$\alpha$ discs with O-abundances seen in old and young planetary nebulae in M~31 respectively, supporting the existence of $\alpha$-bimodality in the inner regions of M~31. Finally, we use globular cluster ages to project the MW and GSE back in time to $z{\sim}3$ and find that their estimated mass, oxygen and iron abundances are strikingly consistent with the mass-metallicity relation of star-forming galaxies at $z{\sim}3$. In the future, increased transparency around the choice of Solar scale and abundance methodology will make combining chemical abundances easier -- contributing to a complete picture of the chemical evolution of all galaxies.
\end{abstract}

\begin{keywords}
stars: abundances -- Galaxy: abundances -- galaxies: high-redshift -- galaxies: abundances -- galaxies: evolution
\end{keywords}

\section{Introduction}
Understanding the emergence and growth of galaxies \emph{and} determining our own galaxy's place in the larger picture, is a major driver for both the fields of extragalactic and Galactic astronomy. To build a complete picture will require marrying knowledge gained from extragalactic studies with discoveries made through Galactic Archaeology \citep{freeman2002}. As the number of increasingly more distant galaxies continue to grow via the James Webb Space Telescope \citep[JWST, e.g., GNz-11 at $z\sim11$, GHZ2 at $z\sim12.3$, JADES-
GS-z14-0 at $z\sim14$ and MoM-z14 at $z\sim14.4$,][]{bunker2023, castellano2024, carniani2024,naidu2025}, and thanks to the ongoing efforts of large surveys using ground-based spectrographs at $z\sim2-3$ \citep[e.g.,KMOS$^{\mathrm{3D}}$, SAMI, SDSS-MANGA, MAGPI, KBSS, VANDELS, MOSDEF;][]{wisnioski2015, bryant2015, bundy2015, foster2021, rudie2012, steidel2014, mclure2018_vandels,kriek2015}, our understanding of the physics of galaxy formation and evolution is rapidly evolving.

In parallel to this, our understanding of the formation history of the Milky Way (MW) is continuing to increase thanks, in-part, to the Gaia mission making 6D phase space information possible for billions of MW stars \citep{gaiadr1, gaiadr2, gaiadr3}, and dedicated spectroscopic surveys charting the chemical evolution of our Galaxy \citep[e.g.,Gaia-ESO, LAMOST, APOGEE and GALAH,][]{gaiaeso, lamost, apogeedr17, galahdr4}. Finally, promising results from chemodynamics of planetary nebulae \citep[PNe,][]{bhattacharya2019a, bhattacharya2019b, bhattacharya2021, bhattacharya2022, arnaboldi2022} and early JWST results in the Andromeda Galaxy \citep[M~31,][]{nidever2024} have opened the door to high precision modeling of our most massive galactic neighbor. Given the rapid development of these three areas of galactic research, the time is now to consolidate what we know about each in the context of \emph{global} galaxy evolution. 

Placing the MW in the context of other galaxies is no easy task, exemplified by the ongoing debate as to the nature of ``MW analogues'' in both observations \citep[e.g., selecting analogues based on structural properties and/or star formation histories,][]{licquia2015, frasermckelvie2019, boardman2020, tan2024} and simulations \citep[e.g., selecting analogues based on total halo mass and/or resemblance to the Local Group,][]{kobayashi2011, wetzel2016, grand2021}. While the overall structure of the MW may be well understood---two disc-like components with distinct vertical and radial scale lengths, an inner spheroid or Bulge, a diffuse outer halo, bar and spiral arms \citep[][for a recent review]{blandhawthorn2016}---important details, and the origin of each, is still the subject of debate \citep[e.g., the length and pattern speed of the bar and transient nature of the spiral arms, see][]{hunt2025}. 

In addition to the MW components having distinct spatial (and velocity) distributions, they also show unique chemical distributions indicative of different star formation histories -- providing critical clues as to the origin and emergence of each \citep[see][for recent reviews of the chemodynamics and formation history of different components]{helmi2020, deason2024}. Beyond different metallicity distribution functions found across the inner and outer halo \citep{gilmore1985, gilmore1995, carollo2007, youakim2020}, thick and thin disc f\citep{hayden2014, hayden2015, sharma2019} and the Bulge \citep{ness2013, johnson2014, ness2016, lucey2021, ardernarentsen2024}, distinctions also exist between the different MW components in $\alpha$-element abundances, \citep[elements associated with the $\alpha$-process e.g., O, Mg, Si, Ca,][]{burbidge1957, kobayashi2020}. This is perhaps most obvious in the $\alpha$-rich inner, and $\alpha$-poor outer halo \citep{wallerstein1962, venn2004, nissen2010, mckenzie2024} and the high-$\alpha$, vertically thick and low-$\alpha$, vertically thin disc \citep{fuhrmann1998, prochaska2000, bensby2003, reddy2006, bensby2014, hayden2015}, where decades of work deriving abundances from stellar spectra has revealed clear distinctions between the coupled components. 

The unique SFHs of each component, implied by the chemical distinctions, have since been confirmed through stellar ages, with the ancient halo (both inner and outer) and high-$\alpha$ disc being significantly older than the low-$\alpha$ disc \citep{sharma2021, xiang2022, queiroz2023}. Despite a strong legacy of study of the MW, many open questions remain: when did each disc emerge, and what did the MW look like prior to disc formation? As the number of galaxies with structural and chemical properties observed at intermediate- and high-$z$ continues to increase, answering these questions through observing increasingly ancient and/or better-resolved MW analogues \emph{and} understanding how common the MW's evolution is in the context of \emph{all} galaxies, will offer an exciting and complementary approach to continued studies of the MW.   

In contrast to the long history of resolved studies in the MW, it has only recently been possible to extract iron abundances of stellar populations directly from spectra of star-forming galaxies that were forming $\gtrsim10$~Gyr ago (at $z\gtrsim2$). Measuring the stellar metallicity (a proxy of the iron abundance) requires high signal-to-noise in the rest-frame far-ultraviolet (FUV) continuum. Early analyses were based on measurements of FUV absorption indices, and were typically restricted to lensed sources \citep[e.g.,][]{rix2004, dessauges_zavadsky2010} or composite spectra \citep[e.g.,][]{halliday2008}, with few measurements of individual sources \citep[e.g.,][]{sommariva2012}. 
More recently, \citet{steidel2016} introduced a more complete methodology wherein the full FUV continuum at $\lambda_{\rm rest} < 2000\,$\AA \ is fit with stellar population synthesis (SPS) models to more robustly constrain the stellar metallicity and iron abundance in comparison to the prior absorption index approach.

A plethora of works have since used this methodology to measure the iron abundances of high-$z$ galaxies with the aim of measuring the stellar mass ($M_{\ast}$)-stellar metallicity ($Z_{\ast}$) relationship \citep[e.g.,][]{cullen2019, chartab2024}. Combining iron abundances with oxygen abundances from rest-frame optical spectra has revealed that high-$z$ star-forming galaxies (SFGs) have supersolar O/Fe ratios, reflecting the lack of enrichment from Type Ia supernovae \citep[e.g.,][]{topping2020a, topping2020b, cullen2021, kashino2022, strom2022, stanton2024}. 
Interestingly, there is preliminary evidence of high-$z$ systems following the same trends as the MW discs in the [O/Fe] vs. [Fe/H] space \citep[e.g.,][]{cullen2021, kashino2022}, based on overlapping positions of high-$z$ galaxies with the chemical evolution models describing MW stars of \citet{kobayashi2020}.
Recent studies \citep{arnaboldi2022, kobayashi2023} have also demonstrated that supernova enrichment processes can be similarly probed in emission nebulae, including star-forming galaxies both locally \citep{bhattacharya2025b} and at high-$z$ \citep{bhattacharya2025,rogers2024,morishita2025,stanton2024b} using O and Ar abundance measurements. In this case, Ar acts as a proxy for Fe based on the contribution of the Fe-peak nucleosynthetic pathway to the production of Ar \citep{kobayashi2020}. 
However, high-$z$ studies of multi-element abundance ratios, such as O/Fe, remain limited by small samples with sufficient signal-to-noise (SNR) to robustly measure Fe, in particular \citep[note that this is changing with JWST, where Fe-measurements out to $z>9$ are possible with $\mathrm{SNR}>40$ spectra][]{ji2024b, nakane2024, nakane2025}. 

Despite these limitations, many interesting abundance trends are emerging in early galaxies as the number of measurements increases. Firstly, it is becoming apparent that enhanced O/Fe abundances are ubiquitous in star-forming galaxies at these redshifts; although independent studies disagree on the typical level of enhancement, all analyses so far find that $\langle \mathrm{[O/Fe]} \rangle > 0$ with high significance \citep[with the exception of GNz-11, which could be dominated by a massive star cluster, explaining an apparent enhancement in N in conjunction with a depletion in O,][]{charbonnel2023, dantona2023, nakane2024}. These findings are consistent with enrichment dominated by short-timescale core-collapse supernovae at these cosmic epochs \citep[e.g.,][]{tinsley1979, matteucci1986, chruslinska2024} and could be the result of high specific star-formation rates (sSFR$=$~SFR$/M_{\ast}$).
Secondly, there is emerging evidence for mass-dependent sequences in [O/Fe] vs. [Fe/H] space suggesting lower star-formation efficiencies and possibly greater outflow efficiencies in lower mass galaxies \citep[e.g.,][]{syblilska2018, vincezo2018, chartab2024, stanton2024, velichko2024}; this trend is consistent with local observations of dwarf galaxies \citep[e.g.,][]{tolstoy2009}.

With the overarching goal of exploring the evolution of galaxies across cosmic time through their chemical abundances and placing the MW and Local Group in the context of what is currently known at Cosmic Noon ($z\sim2-3$), we assembled the ChemZz\footnote{Upper case ``Z'' in this context refers to the total abundance of metals (see Sec.\ref{sec:toolkit}, while lower case ``z'' refers to redshift.} collaboration of chemical abundance enthusiasts. In this study, we present our first attempt to homogenize extragalactic and Galactic chemical abundance datasets, revealing similarities and differences between the two. In this first paper, we focus only on the evolution of oxygen as a function of iron, as it is well-measured in both environments. To do this, we explore abundances in populations of old, evolved stars in the MW and the Local Group assuming that they act as proxies of the young stellar populations observed in our $z\sim2-3$ sample.

The paper is organized as follows, we begin by introducing the ``Toolkit'' in Sec.~\ref{sec:toolkit}, where we attempt to bring readers from both the extragalactic and Galactic community ``up to speed'' on field-specific terminology and methodology. In Section~\ref{sec:mwabunds}, we discuss our compilation of MW and Local Group abundances, define our selection of different MW structural components, describe how we bring all of the abundances onto a common scale \citep[that of][]{kobayashi2020} and include a discussion of systematics. In Sec.~\ref{sec:highzcomp}, we discuss our sample of Cosmic Noon abundance measurements and again include a discussion of systematic uncertainties. In Sec.~\ref{sec:abundpatt}, we present our homogenized dataset of abundances in chemical space, interpret trends and compare and contrast abundances seen in our Cosmic Noon and Local galaxy populations. In Sec.~\ref{sec:chronotrends}, we explore incorporating ages for MW field stars and globular clusters in an effort to place the MW at $z\sim3$ or a look-back time of 11.7~Gyr. We conclude by highlighting a first attempt to assess the location of the MW on the $z\sim3$ mass-metallicity relation for star-forming galaxies. Finally, we provide a discussion of next steps and a ``wish list'' for both the extragalactic and galactic communities to consider as we pursue a complete picture of galaxy evolution together.

\begin{table*}
\centering
\caption{Overview of the abundance samples used in this study, highlighting the survey or object of focus and the relevant reference. In the case of M~31, we adopt the binned values for the inner planetary nebulae found at radial distances of between 3-14~kpc. As defined previously, ``SFG'' refers to star forming galaxies.  \label{tab:deluxesplit}  }
\begin{tabular}{lcccll}
\hline
Survey/Object & Redshift ($\langle z\rangle$) & Iron ($N$) & Oxygen ($N$) & Reference(s) & Notes \\
\hline
\multicolumn{6}{c}{Local Group} \\
\hline
Milky Way, GSE & 0 & 78691, 1557 & 78691, 1557 & GALAH DR3: \citet{galahdr3} & Stellar abundances \\
M~31 & 0 & 10  & 10  & \citet{bhattacharya2022} & Planetary nebulae  \\
Globular Clusters & 0 & 16 & 16 & \citet{carretta2010} & Stellar abundances \\
LMC & 0 & 17  & 17  & \citet{pompeia2008} & Stellar abundances  \\   
    & & 99 & 99 & \citet{vanderbarlmc} & Stellar abundances \\
SMC & 0 & 165  & 165  & \citet{mucciarelli2023} & Stellar abundances  \\
Sagittarius, M~54 & 0 & 27, 76  & 27, 76  & \citet{carretta2010} & Stellar abundances  \\

\hline
\multicolumn{6}{c}{Cosmic Noon} \\
\hline
NIRVANDELS & 3.5 & 17 & 34 & \citet{cullen2021, stanton2024} & Individual SFGs \\
 & 3.5 & 4 & 4 & \citet{cullen2021, stanton2024} & Stacked SFGs \\
KBSS & 2.4 & 1 & 1 & \citet{steidel2016} & Stacked SFGs \\
 & 2.3 & --- & --- & \citet{strom2022} & MZR locus \\
LATIS & 2.5 & 1 & 1 & \citet{chartab2024} & Stacked SFGs \\
\hline
\end{tabular}
\end{table*}

\section{The ``Toolkit''}
\label{sec:toolkit}
In this section, we introduce the terminology and concepts referenced throughout the remainder of the paper, with the goal of placing readers from both the Galactic and extragalactic communities on equal footing. We begin by defining ``metallicity'' and ``abundance,'' transition to discussing abundance scales (including our preferred scale), and finish by reviewing the connection between the position of systems in abundance space and galactic chemical evolution.

\subsection{Terminology}
\label{subsec:term}
There are many ways of referring to the enrichment of heavy elements (beyond He) in astrophysical systems. The vocabulary used in a given setting depends on the object(s) being studied, the method(s) employed to study them, and convention---i.e., the choice is partly physical and partly practical. The most basic (if deceptively simple) concept is that of ``metallicity'' $Z$, which refers to the \emph{mass fraction} of all elements heavier than helium; $Z$ is often reported relative to the solar metallicity, $Z_{\odot}\sim0.014$ \citep{asplund2009}, and may specifically refer to the metallicity of stars ($Z_{\ast}$) or the metallicity in the gas \citep[$Z_{\textrm{gas}}$ or, sometimes, $Z_{\textrm{neb}}$; see ][for a more detailed description]{maiolino2019}. Note that the value of $Z_{\odot}$ may vary slightly across different fields in astrophysics and is revised periodically as our knowledge of the Sun has improved \citep[e.g., most recently in][]{asplund2020}.

However, in \emph{practice}, astronomers do not measure $Z$ observationally but instead measure the amount of specific elements. So, it is more common to refer to the ``abundance'' of these elements, defined as A(X)~$=12+\log_{10}(\textrm{X/H})$, where X/H stands in for the \emph{number} (or number density) of atoms of some element relative to hydrogen and the abundance of hydrogen is defined as being equal to 12. Abundances are also frequently reported relative to solar abundances, usually using the so-called bracket notation, so that [X/H]~$=\log_{10}(\textrm{X/H})-\log_{10}(\textrm{X/H})_\odot$; both A(X) and square bracket notation were introduced by \citet{helfer1959}. The relative abundance of different heavy elements can be reported using the same notation, so the ratio of oxygen (O) to iron (Fe) might be written as [O/Fe]~$=\log_{10}(\textrm{O/Fe})-\log_{10}(\textrm{O/Fe})_\odot$. For a recent review of abundance determination techniques and applications in stellar spectra, see the work by \citet{nissen2018}.

As a shorthand, many studies use the word ``metallicity'' when referring to A(X); for example, this is very common in studies of star-forming galaxies, where oxygen is usually the easiest (and sometimes only) elemental abundance that can be measured. This substitution has the practical benefit of allowing direct comparison with theoretical models (which often use $Z$) but also has the obvious drawback of implicitly assuming that the system under consideration has a \emph{solar-scaled} abundance pattern. However, [X/H]~$=-1$ only corresponds to $Z=0.1~Z_{\odot}$ if \emph{all} elements have [X/H]~$=-1$. Yet, in many cases, the relative abundance of different elements is not solar (i.e., [X/H] is not equal for all elements or [X$_2$/X$_1$] is nonzero), and these differences can have important physical implications (see Section~\ref{subsec:tinsley}). Consequently, eliding the distinction between ``abundance'' and ``metallicity'' may introduce confusion and uncertainty unless a study is very clear about their assumptions.

Bearing all of this in mind, in this study we will only use the term ``metallicity'' when referring to the total iron abundance ([Fe/H]) in each system, as is the convention in studies of resolved stellar abundances in the Local Group \citep[see e.g.,][for large abundance compilations]{venn2004, tolstoy2009}. All Milky Way and Local Group abundances used throughout this study (with the exception of M~31) are measured from  resolved stellar spectra of individual stars. More specifically, [Fe/H] is determined from absorption features in the stellar spectra associated with different atomic transitions occurring in the stellar atmosphere. Note that [Fe/H] may be derived from measurements of singly-ionised iron or doubly-ionised iron, or an average of the two \citep[again, see][for a review of the techniques used to determine stellar abundances]{nissen2018}. Measurements of metallicity in our high-$z$ sample (often referred to as ``stellar metallicities'' in that community) are determined using various methods, although most operate on a similar principle to the method used to measure individual stellar metallicities, just for a \emph{population} of stars. 

In addition to Fe, we also discuss O throughout this study. As for Fe, A(O) as reported in the various Local Group studies is determined from O absorption lines present in the stellar spectra (again, with the exception of M~31, see Section~\ref{subsec:localgroup_abund}). For the Local Group, both [Fe/H] and A(O) (or [O/H]) refer to \emph{stellar} enrichment, as the abundances of these elements are determined directly from stellar spectra; the same is true for high-$z$ measurements of [Fe/H]. However, in contrast, A(O) in high-$z$ galaxies is measured using emission lines originating from the hot, ionised gas surrounding massive stars in star-forming regions and is often described as the ``gas-phase metallicity.'' For the sake of clarity and to distinguish it from A(Fe) and [Fe/H], we will not refer to A(O) or [O/H] as ``metallicity'' in this work, even though it is common parlance in studies of star-forming galaxies beyond the Local Group.

\subsection{Abundance Scales}
\label{subsec:abundscales}
All of the data sets compiled in this work have chosen to place themselves on \emph{a} particular Solar scale. Despite adopting a common reference object (the Sun), values for the absolute Solar abundance of Fe and O can vary from study to study by magnitudes comparable to the abundance uncertainties \citep[e.g., 0.08~dex in the case of O derived using two different treatments of 3D NLTE,][]{steffan2015, asplund2009, asplund2020}. To compare abundance patterns across our compilation of MW and high-$z$ data, we must first reconcile the various choices of abundance scale and zero point.  

To overcome these differences, we place all of the Fe and O abundances in our compilation onto the Solar scale of \citet[][hereafter K20]{kobayashi2020}, who in turn adopt A(Fe)$_\odot=7.50$ from \citet{asplund2009} and A(O)$_\odot=8.76$ from \cite{steffan2015} for a bulk (proto-)solar metallicity of $Z_{\odot}=0.0144$; for completeness, \citet{asplund2009} report a bulk (proto-)solar metallicity of $Z_\odot=0.0142$. Note that we will specify the choice of Solar scale used throughout when discussing or plotting [O/Fe] or [Fe/H]. For example, values on the ``K20 scale'' are denoted as $\mathrm{[Fe/H]_{K20}}$ and $\mathrm{[O/Fe]_{K20}}$.

\subsection{The Tinsley-Wallerstein Diagram \& Galactic Chemical Evolution}
\label{subsec:tinsley}
\begin{figure}
    \begin{centering}
    \includegraphics[width=\linewidth]{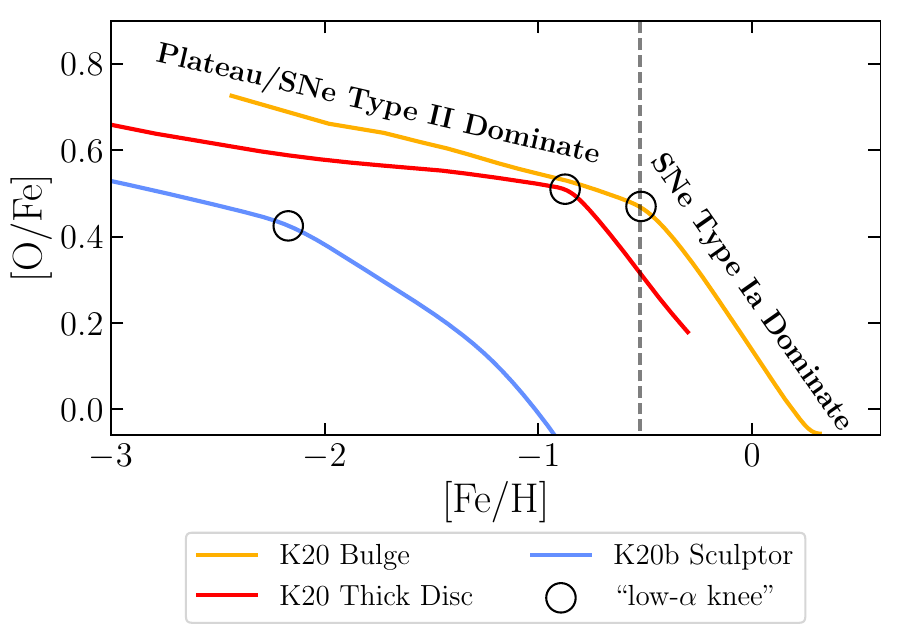}
    \caption{Representation of the ``Tinsley-Wallerstein'' diagram \citep{wallerstein1962, tinsley1979}, depicting the predicted chemical evolution of three different systems as a result of different star formation histories. Nucleosynthetic sources (driving the evolution) are marked, as well as the location of the ``low-$\alpha$'' knee in the three systems. Predictions for the evolution of the MW bulge and high-$\alpha$ disc modeled by \citet{kobayashi2020} and the Sculptor dwarf galaxy modeled by \citet{kobayashi2020b} are marked in yellow, red and blue, respectively.}
    \label{fig:tinsley}
    \end{centering}
\end{figure}

To interpret the set of chemical abundances and connect them to galaxy evolution, we use the Tinsley-Wallerstein diagram \citep{wallerstein1962, tinsley1979}. This diagram (depicted in Fig.~\ref{fig:tinsley}) traces formation timescales by comparing elements originating from different nucleosynthetic sites. On the $y$-axis (ordinate), an element (or an average of several elements) associated with the $\alpha$-process 
is plotted as a function of metallicity ([Fe/H], see Section~\ref{subsec:term}) marked on the $x$-axis (abscissa). The dominant fusion site for $\alpha$-elements is inside massive stars (through hydrostatic burning) and through the subsequent core-collapse \citep[explosive burning, in Type II, Ib, and Ic supernova;][]{woosley1995, woosley2002, kobayashi2006, kobayashi2020}. As a result, the timescales for $\alpha$-element production are relatively short, commensurate with the lifetime of massive stars, meaning that elements like O and Mg re-enter the interstellar medium (ISM) on timescales of  $10-100$~Myr.

In addition to the $\alpha$-elements, massive stars also introduce iron into the ISM at early times. This balance of both $\alpha$-element and iron production produces a plateau in [$\alpha$/Fe] over a range in metallicity, as seen in Fig.~\ref{fig:tinsley} (where we have chosen O as our $\alpha$-element tracer). After as little as $\sim100$~Myr, low-mass stars end their lives as Type~Ia supernovae (SNe), synthesizing and releasing Fe-peak elements (e.g., Ti, Cr, Mn, Fe, Ni, Cu) into the ISM \citep{woosley2002, kobayashi2006}. Because Type~Ia SNe produce Fe but not $\alpha$-elements, the overall ratio of [$\alpha$/Fe] declines as a function of increasing metallicity. The onset of Type~Ia SNe is associated with the appearance of the ``low-$\alpha$ knee'' in the Tinsley-Wallerstein diagram---providing a timestamp in metallicity for a galaxy's evolution (though \citealp{mason2024} find that the appearance of a ``low-$\alpha$ knee'' in simulations is only associated with the onset of Type Ia \emph{if} a galaxy has a steadily declining star-formation rate, we revisit this later).

Connecting the Tinsley-Wallerstein diagram to galaxy evolution, $\alpha$-element production and entry into the ISM is sensitive to both the number of massive stars as well as the star-formation rate (SFR), or equally, the star formation efficiency (SFE), if the efficiency is assumed to be constant as in the case where $N=1$ in the Schmidt-Kennicutt star formation law such that $\mathrm{SFR}=\mathrm{SFE}\times\mathrm{M}_{\mathrm{gas}}$ \citep{schmidt1959, kennicutt1998}. As a result, galaxies that have high star formation efficiencies, or equally high sSFRs, \emph{and} are continuously forming stars, maintain a plateau in [$\alpha$/Fe] to higher values of [Fe/H], than galaxies that are less efficient or ``slow'' at forming stars, even after the onset of Type~Ia enrichment. Examples of this are shown in Fig.~\ref{fig:tinsley} where the rapid assembly of the MW Bulge (represented with a chemical evolution model in yellow from \citealp{kobayashi2020}) appears distinct from the slow star-formation history (SFH) of the much less massive Sculptor dwarf galaxy (dGal, in blue). The positions of the low-$\alpha$-knees in these two galaxies are significantly displaced in [Fe/H], with SF halting in Sculptor at significantly lower metallicities than the Bulge. The \citet{kobayashi2020} model for the MW high-$\alpha$ disc is also included in Fig.~\ref{fig:tinsley} in red, highlighting another system that experienced efficient star formation. 

In summary, the SFH and temporal evolution of a galaxy are reflected in their evolution through the Tinsley-Wallerstein diagram. To derive a SFH from the chemical evolution of a galaxy through the diagram, analytical galactic chemical evolution (GCE) models are used, with the simplest being the ``closed-box model.'' In the closed-box model, enriched gas is converted into stars with some global star formation efficiency (SFE) and no loss or gain to the mass budget via outflows or inflows. More modern GCE models \citep[see][for a full review]{matteucci2021} involve more realistic galactic physics \citep[including inflows and outflows, e.g., OMEGA,][]{omega}, bursty star formation \citep[e.g., flexCE,][]{flexce}, stellar migration \citep[e.g., VICE,][]{vice} and multiple nucleosynthetic channels tracing both common and rare events \citep[e.g., neutron star mergers, magneto-rotational supernovae,][]{kobayashi2020}. Throughout this paper, we will use both observational data and GCE models from \citetalias{kobayashi2020} to interpret the chemical evolution of the galaxies in our sample.

\section{Milky Way \& the Local Group}

\label{sec:mwabunds}
In this section we describe the selection criteria we apply to the MW field star data and our chemodynamical definitions of the MW structural components, namely: the low-$\alpha$ (thin) and high-$\alpha$ (thick) discs, and the inner and outer halo. We also discuss the data sets we include for MW globular clusters and Local Group galaxies. The section concludes with a discussion of our efforts to bring the various samples onto a common abundance scale and some systematic uncertainties relevant for studies of resolved stellar abundances.

\subsection{GALAH DR3 Sample Selection}
Oxygen abundance measurements of resolved MW field stars are taken from the third data release of the GALactic
Archaeology with HERMES (GALAH) survey \citep{galah, galahdr3}. GALAH is a medium resolution (R~$\sim28,000$) spectroscopic survey of 588571 Southern hemisphere stars (81.2\% of which are within 2~kpc of the Sun) conducted using the HERMES spectrograph \citep{hermes} and the 2dF fibre positioning system \citep{2df} on the 3.9~m Anglo-Australian Telescope. The wavelength coverage (although discontinuous) spans $4713-7887$~\AA, capturing absorption features associated with 30 unique elements, across four nucleosynthetic channels ($\alpha$, Fe-peak, $s$- and $r$-process). Very broadly, stellar abundances in DR3 are determined using Spectroscopy Made Easy \citep{valenti1996, piskunov2017}, a spectrum synthesis code and 1D MARCS model atmospheres \citep{marcs}, with the final stellar abundances determined through $\chi^{2}$ minimisation of the synthetic and observed spectra on an element-by-element basis \citep[see Section 3.2 in][for a more detailed discussion]{galahdr3}. We adopt DR3 for this study, rather than DR4, because the abundances are derived using a standard approach for which the systematics are better understood at present \citep[vs. abundances derived using a machine learning approach, as in DR4;][]{galahdr4}.  

We choose to adopt GALAH oxygen abundances because they are determined using optical \ion{O}{I} absorption lines\footnote{specifically the three \ion{O}{I} near-IR lines at 7771.94\AA, 7774.17\AA\, and 7775.35\AA} and have been corrected to account for deviations from local thermodynamic equilibrium (NLTE corrections) following \citet{amarsi2020}. This is important as NLTE is implicitly assumed as part of the abundance derivations in our Cosmic Noon dataset. Unfortunately, O abundances derived from IR spectra from the Apache Point Observatory Galactic Evolution Experiment \citep[APOGEE, ][]{apogeedr17} have not been NLTE corrected, likely because they are measured from OH molecular lines \citep{jonsson20}, so we neglect them in this study. Furthermore, systematic differences are known to exist between O abundances derived from optical spectra with those derived from infrared spectra, complicating efforts to combine data from different surveys \citep{carrillo2022}.  

To construct our catalogue of GALAH DR3 stars, we follow \citet{monty2024} and the recommendations from GALAH regarding which flagged stars to remove.\footnote{\url{https://www.galah-survey.org/dr3/using_the_data/}} That is, we remove stars with \texttt{snr\_c3\_iraf}~$<30$, \texttt{flag\_sp}~$\neq0$, \texttt{flag\_fe\_h}~$\neq0$ and \texttt{flag\_O\_fe}~$\neq0$. To sample the old stellar populations in the MW---which are most analogous to the young, massive stellar populations observed in our Cosmic Noon sample---we select only red giant branch stars in our study, stipulating cuts on the effective temperature (T$_{\mathrm{eff}}\leq5300$~K) and surface gravity (log~$g\leq3$). Finally, we require abundance uncertainties of less than 0.2~dex in [O/Fe] and [Fe/H] to build our clean dataset. In total, we retain $\sim80,000$ stars from the original DR3 catalogue, with an average T$_{\mathrm{eff}}\sim4650$~K and log~$g\sim2$.

\subsection{Defining Milky Way Components}
\label{sec:mwcomp}
Under the assumption that the different structural components of the MW emerged at different times, we aim to explore the evolution of each in relation to our Cosmic Noon samples with fixed look back times. To do this, we split the GALAH data into four structural components: the low-$\alpha$ (thin) and high-$\alpha$ (thick) discs and inner and outer halo. Because these components were formed at different times via different channels, they have distinct chemodynamical signatures. Consequently, we can apply various dynamical and chemical cuts to isolate each component, as described below, relying on both historical definitions and recent discoveries. 

Dynamical properties for the GALAH DR3 stars are taken from the \texttt{GALAH\_DR3\_VAC\_dynamics\_v2} value-added catalogue described in \citet{buder2022}. To calculate $E$, $L_{z}$ and other orbital parameters, \citet{buder2022} adopt the MW potential from \citet{mcmillan2017} implemented within the \texttt{galpy} galactic dynamics package \citep{bovy2015}. \citet{buder2022} adopt a solar radius of 8.21~kpc and orient the Sun 25~pc above the plane, following \citet{juric2008}. Their total solar velocity of ($U, V, W$) = (11.1, 248.27, 7.25)~km/s in keeping with \citet{shonrich2010}, after assuming the same circular velocity at the Sun (233.1~km/s) as \citet{mcmillan2017}.  

\begin{figure*}
    \centering
    \includegraphics[scale=0.4]{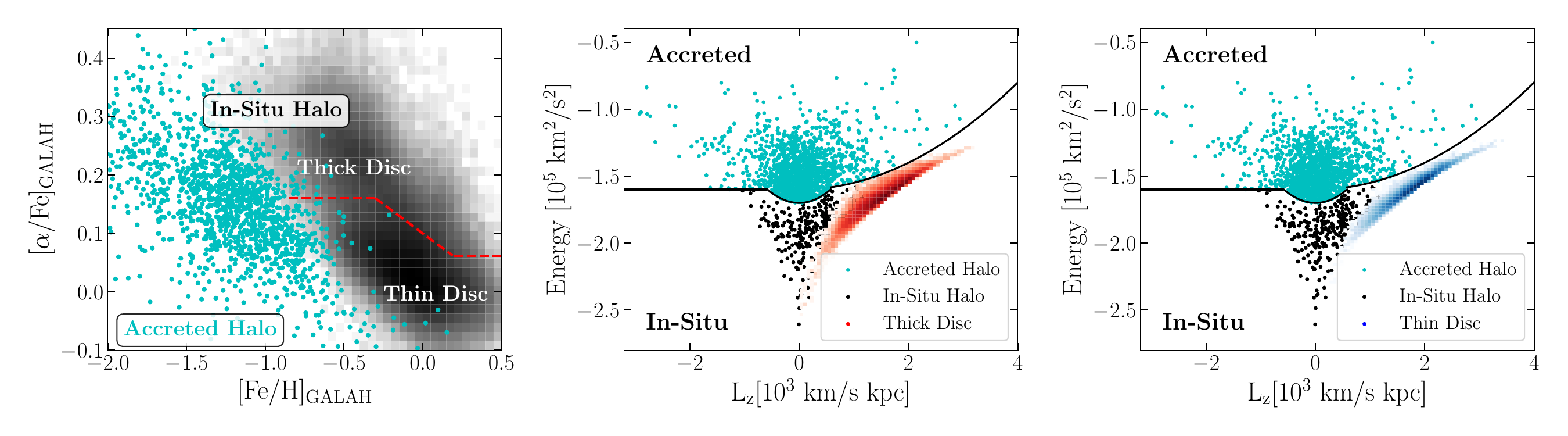}
    \caption{The four MW components from our GALAH sample shown in both chemical and dynamical space. From left to right, the first column shows our definition for the four components in [$\alpha$/Fe] vs. [Fe/H] space. The three in-situ components are shown in black, while stars from the accreted halo are marked in cyan. Our chemical definition separating the high-$\alpha$ (thick) and low-$\alpha$ (thin) discs is marked in red, following the methodology of \citet{mackereth2019}. The second column shows the location of the accreted (cyan) and in-situ (black) halo components in energy ($E$) vs. z-component of angular momentum ($L_{z}$) separated by the boundary derived in \citep{belokurov2023}. The location of the high-$\alpha$ disc is marked in red. The third column repeats what is shown in the second, with the exception that the low-$\alpha$ disc is now shown in blue.}
    \label{fig:galah_MW_comp}
\end{figure*}

\subsubsection{In-Situ \& Accreted Halo}
The first cut we make to isolate MW components is to define \emph{in-situ} and \emph{accreted} stars within the MW. We consider the more spatially compact inner MW halo \citep[also referred to as the high-$\alpha$ halo or ``Aurora'',][]{nissen2010, bonaca2020, Belokurov+2022} and the low-$\alpha$ and high-$\alpha$ discs to be the predominant in-situ components. Before applying chemical cuts, we first split both the GALAH sample into in-situ and accreted stars based on energy, $E$ and $z$-component of the angular momentum, $L_{z}$. We apply the $E-L_{z}$ boundary derived in \citet{belokurov2023} motivated chemically by a difference in [Al/Fe] seen between the inner and outer Galaxy, and assumed to be due to different SFHs---with Al acting as a tracer of massive stellar evolution. As in \citet{monty2024}, the exact $E-L_{z}$ boundary is adjusted to accommodate the different choice of Galactic potential between \citet{belokurov2023} and \citet{buder2022}. After making the adjustment, we require in-situ stars in GALAH to satisfy the following, where $L_{z}$ is in units of [km/s kpc] and $E$ is units of [km$^{2}$/s$^{2}$]:
\begin{align*}
    L_{z}/10^{3} < -0.58&: E/10^{5} < -1.6 \\
    -0.58 \leq L_{z}/10^{3} < 0.58 &: E/10^{5} < -1.7 + 0.3L_{z}^{2}\\
    L_{z}/10^{3} > 0.58 &: E/10^{5} < -1.6 + 0.5L_{z}^{2}
\end{align*}

After applying cuts in $E, L_{z}$, chemical cuts are applied to remove stars with $\mathrm{[Al/Fe]_{GALAH}} \leq -0.1$ below metallicities of $\mathrm{[Fe/H]_{GALAH}} = -0.6$, as these are likely accreted interlopers. Finally, in-situ stars are defined as having $\mathrm{[Mg/Fe]_{GALAH}} \leq -0.3\times\mathrm{[Fe/H]_{GALAH}}$. Note that the bulk of these cuts are made to exclude stars below the cut-off in $E, L_{z}$ that likely belong to the Gaia-Sausage/Enceladus (GSE) dwarf galaxy \citep{haywood2018, helmi2018, belokurov2018}---the likeliest candidate for the last major merger experienced by the MW. 
We assign stars which sit above the boundary in $E, L_{z}$ without applying chemical cuts as accreted halo/GSE stars in GALAH. 
These stars are marked with the cyan points in Fig.~\ref{fig:galah_MW_comp}.  The middle and right panel of Fig.~\ref{fig:galah_MW_comp} show the distributions of the two populations in $E, L_{z}$-space, with the black stars belonging to the in-situ halo and the blue, the accreted halo.

\subsubsection{High-$\alpha$ \& Low-$\alpha$ Discs}
The origin of the spatial bimodality of the MW low-$\alpha$ (thin) and high-$\alpha$ (thick) discs is still uncertain. Although it is generally accepted that the thick disc is older(supported by its high $\alpha$-abundance), the source of its increased vertical scale height is debated---e.g., perhaps it arises through dynamical heating via the last major merger \citep{mackereth2019} or perhaps it was born with a large vertical velocity dispersion due to turbulent gas \citep{wisnioski2015}. Additionally, the origin of the light- ($\alpha$-)element bi-modality between the two discs and its relation to the spatial differences between the discs as well as its prevalence in other galaxies, also remains uncertain (we discuss this further in Sec.~\ref{sec:mwm31}). Locally (in the Solar neighborhood), it has been suggested that the $\alpha$-bimodality could be the result of radial migration, potentially related to the growth and deceleration of the MW bar \citep{schonrich2009, minchev2013, sharma2021, zhang2025}. Regardless of its origin, we split the discs primarily based on this $\alpha$-bi-modality and verify our results using the spatial distributions.

We make two dynamical cuts to select disc stars among our in-situ sample, namely that $|z_{\mathrm{max}}|<3$~kpc and $v_{\phi} >= 120$~km/s to ensure that the stars are both i) confined to the disc of the Galaxy, and ii) display coherent rotation. After these cuts we follow the methodology of \citet{mackereth2019} to split the high$-\alpha$ and low-$\alpha$ discs based on the evolution of their [$\alpha$/Fe] abundance as a function of metallicity. Following \citet{mackereth2019}, we separate the GALAH in-situ samples visually using their 2D histograms in [$\alpha$/Fe]$_\mathrm{GALAH}$ vs. [Fe/H]$_\mathrm{GALAH}$ space, adopting the bulk [$\alpha$/Fe] values reported by GALAH. To make a clearer distinction between the two disc populations, we also restrict the low-$\alpha$ population to metallicities above $\mathrm{[Fe/H]=-0.6}$. The separation we adopt is shown as the red dashed line in Fig.~\ref{fig:galah_MW_comp}, with the two components annotated. The high-$\alpha$ and low-$\alpha$ disc sample distributions in $E, L_{z}$-space are also shown in the middle and right panel of Fig.~\ref{fig:galah_MW_comp} as the red and blue density distributions, respectively. Because we rely primarily on chemical cuts to separate the thick and thin discs, moving forward we refer to these two components solely as the high-$\alpha$ and low-$\alpha$ discs, respectively.

\subsubsection{Shifting onto the K20 Abundance Scale}

To shift our GALAH DR3 sample onto our preferred scale (K20), we bring the individual [O/Fe]$_\mathrm{GALAH}$ and [Fe/H]$_\mathrm{GALAH}$ measurements back to A(O) and A(Fe) assuming the Solar values published in \citet{galahdr3} (A(O)$_\odot$ = 8.77, A(Fe)$_\odot$ = 7.38), before applying the K20 A(O) and A(Fe) values. \citet{galahdr3} derive their Solar abundance values through measurements of the Solar spectrum observed with the same instrumental set-up as the GALAH survey and verify their scale through reproducing Solar-like abundance patterns in Sun-like stars in the Solar neighbourhood. 

\subsection{\label{sec:gcabunds}Globular Clusters}
To supplement our field star population and provide accurate ages for older stars, we consider [O/Fe] and [Fe/H] abundances for 16 MW globular clusters (GCs) from \citet{carretta2010}. Due to their pronounced chemical complexity \citep[see, e.g.,][for a recent review]{milone2022}, computing a single mean value of [O/Fe], and in some cases [Fe/H], for the set of MW GCs is non-trivial; as more and more GCs are being found to host statistically significant spreads in metallicity \citep[see, e.g.,][]{mckenzie2022, legnardi2022, monty2023a}. In the case of [O/Fe], a large spread in O driven by the appearance of an O-normal and O-depleted population is an example of one of the multiple anti-correlations found among light elements in nearly all GCs and produced via the CNO cycle and proton capture chain \citep[resulting in an O-depletion in conjunction with an Na-enhancement,][]{gratton2004}. Despite this complexity, the mean [$\alpha$/Fe] abundances in GCs have been shown to agree with their dynamical origins, with accreted GCs showing lower [$\alpha$/Fe] abundances than in-situ clusters at the same metallicity in agreement with field star classifications and reflecting a difference in star formation efficiency between disrupted dwarf galaxies and the MW \citep{monty2023b, belokurov2023, belokurov2024, monty2024, ceccarelli2024}. 

Under the assumption that the O-normal population of stars in GCs sample the oxygen abundance of their host galaxies at the time of GC formation, we determine [O/Fe] and [Fe/H] for the 17 GCs in the \citet{carretta2009} sample \emph{from these stars alone} and shift them onto our preferred scale (K20). The original \citet{carretta2009} data assumes the Solar scale published in \citet{gratton2003}, namely that A(Fe)$_{\odot}=7.54$ and A(O)$_{\odot}=8.79$. Therefore, we added these values back to the $\mathrm{[O/Fe]_{CAR}}$ and $\mathrm{[Fe/H]_{CAR}}$ values published by \citet{carretta2009} before subtracting the K20 Solar A(O) and A(Fe) values. Note also that \ion{O}{I} abundances were determined from the two forbidden lines at 6300\AA\, and 6363\AA, which are unaffected by NLTE effects \citep{amarsi2019}. To determine the abundance of [O/Fe]$_{\mathrm{GC}}$ and [Fe/H]$_{\mathrm{GC}}$ for each GC, we average only among the first generation stars with $\mathrm{[O/Fe]_{CAR}}\geq\mathrm{[O/Fe]_{ave, _{CAR}}}-\sigma\mathrm{[O/Fe]_{_{CAR}}}$ (where $\sigma$ is the standard deviation in this case) and 
among stars with $\mathrm{T_{eff}}\leq5300$~K, log~$g\leq3$ and [X/Fe] uncertainty less than 0.2~dex, in keeping with our GALAH sample selection criteria. On average, these cuts result in mean cluster abundances being driven by stars with $\mathrm{T_{eff}}\sim4400$~K and log~$g\sim1.5$. 

Following the classifications from \citet{belokurov2023}, 10/16 GCs in our sample are classified as having formed in-situ (i.e., alongside the primordial MW) and 6/16 are classified as being accreted \citep{belokurov2024}. Of the six accreted GCs, we re-classify NGC~288 as an in-situ cluster following the findings of \citet{horta2020} and \citet{monty2023b} and assign three of the remaining five GCs as originating in the GSE progenitor galaxy following the classifications of \citet{myeong2019} and \citet{massari2019}. We revisit the implication of these classifications in Section~\ref{sec:chronotrends}.

\begin{figure*}
    \centering
    \includegraphics[scale=0.48]{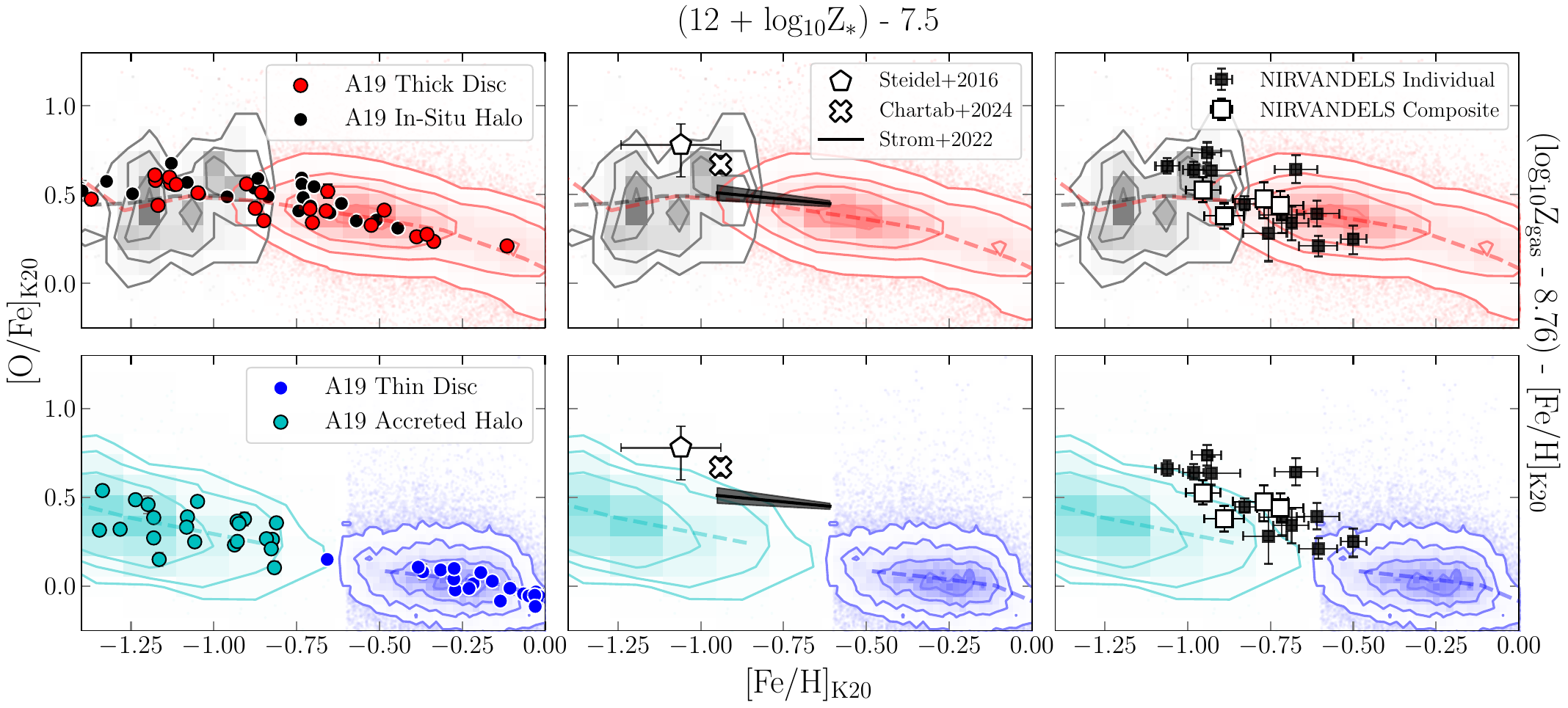}
    \caption{Density contours of the four MW structural components we consider in [O/Fe] vs. [Fe/H] space using our cleaned GALAH catalogues after shifting into agreement with components in \citetalias{amarsi2019}. \emph{Left,} high resolution, 1D NLTE data for each component taken from the study of \citet{amarsi2019} is over-plotted as the colour-coded points using their dynamical associations. The mean value of each component as a function of metallicity in GALAH DR3 is marked using a dashed line. Both GALAH and \citet{amarsi2019} have been put onto the Solar scale adopted by \citet{kobayashi2020} (''K20''). \emph{Middle,} stacked data from the KBSS \citep{steidel2014} and LATIS surveys \citep{chartab2024} and the KBSS (oxygen, iron) mass-metallicity relation from \citep{strom2022} at $z\sim2$ (shown in black, with the uncertainty in the relation shown as the shaded region). \emph{Right,} the NIRVANDELS data ($z\sim3$) from \citet{stanton2024} for individual galaxies and composite data (stacked) is included as white squares, shifted to the K20 scale. All intermediate redshift data points have been corrected for dust depletion. \textbf{Note the excellent agreement between the GALAH DR3 abundances across metallicities and the A19 data points in each component after applying a global shift in $\mathrm{[O/Fe]_{K20}}$ and the apparent overlap  between the intermediate redshift data points and the high-$\alpha$ populations in the MW.}}
    \label{fig:galah_amarsi}
\end{figure*}

\subsection{Abundances in the Local Group}
\label{subsec:localgroup_abund}
To explore Cosmic Noon galaxy abundances in the context of as many Local Group galaxies with different stellar masses and SFHs as possible, we also include data from five additional studies. Data for 116 stars in the Large M. Cloud\footnote{We acknowledge that the continued use of the name Magellan is both potentially traumatic for Indigenous peoples and factually incorrect, as he did not discover the clouds \citep{delosreyes24}. We abbreviate the name for the remainder of this paper.} come from the combined studies of \citet{pompeia2008} and \citet{vanderbarlmc} examining the LMC disc and bar respectively. Data for 206 stars from the Small M. Cloud (SMC) are taken from \citet{mucciarelli2023}, while data for both the main body of the Sagittarius dwarf galaxy (Sgr, 27 stars) and its nuclear star cluster (M~54, 76 stars) are taken from \citet{carretta2010}. 

In addition to local dwarf galaxies, we also include data collected in M~31 from the study of \citet{bhattacharya2022} by converting their published O and Ar abundances to [O/Fe]${_\mathrm{K20}}$ and [Fe/H]${_\mathrm{K20}}$. Using spectroscopic observations of planetary nebulae (PNe) identified throughout the disc of M~31 \citep{bhattacharya2019a,bhattacharya2021}, \citet{bhattacharya2022} directly measure the abundance of O and Ar from highly-ionised emission lines. In addition to abundances, the PNe have ages based on extinction, splitting the sample into old ($>4.5$~Gyr) and young ($<2.5$~Gyr) PNe \citep{bhattacharya2019b}. The O and Ar abundances in both PNe populations were converted to [O/Fe]$_{\mathrm{K20}}$ and [Fe/H]$_{\mathrm{K20}}$ using GCE models as described in \citet{kobayashi2023}, noting that a substantial fraction of Ar is synthesised alongside Fe in Type Ia SNe (allowing it to act as a tracer of Fe-peak nucleosynthesis). We adopt the binned [O/Fe]$_{\mathrm{K20}}$ and [Fe/H]$_{\mathrm{K20}}$ data for the two PNe populations in the radial distance range of 3-14~kpc presented in \citet[][see also \citealp{arnaboldi2022}]{kobayashi2023}. As stated in \citet{kobayashi2023}, uncertainties on the binned PNe [O/Fe]$_{\mathrm{K20}}$ and [Fe/H]$_{\mathrm{K20}}$ values are taken from the GCE-transformed measurement error added to the binned standard deviation in quadrature and the bin-width, respectively.

We place all of the additional data sets onto our preferred solar scale thanks to each study explicitly stating their choice of solar scale. That is, A(O) = 8.83, A(Fe) = 7.50 in the case of the LMC \citep{pompeia2008, vanderbarlmc}, A(O) = 8.76, A(Fe) = 7.50 in the case of the SMC \citep{mucciarelli2023} and the K20 abundances in the case of M~31 \citep{bhattacharya2022}. Regarding the choice and/or impact of NLTE corrections in each study, as in the case of the globular cluster data, all three studies we consider utilise either one or two forbidden \ion{O}{I} lines at 6300\AA\, and 6363\AA\ to measure O, both of which are unaffected by NLTE effects \citep{amarsi2019}. Given the range of stellar parameters across all three studies (T$_{\mathrm{eff}}\sim4100$~K, log~$g\sim1.0$ on average), NLTE corrections to Fe are negligible \citep{lind2012}. Finally, we do not apply NLTE corrections to the M~31 data, as the abundances were determined directly from emission lines.

\subsection{\label{sec:local_systematics}Towards a Common Abundances Scale \& Systematic Uncertainties I: Local Sample}
GALAH is a massive survey, and deriving reliable stellar abundances across such a broad range of stellar parameters and data quality is a mammoth task. To validate the reliability of the GALAH DR3 [O/Fe] values in our selected subset across both metallicity and MW components, we compare the mean GALAH abundances to those derived by \citet[][hereafter A19]{amarsi2019}. In their study, \citetalias{amarsi2019} re-analyze high resolution, high signal-to-noise spectra for stars in the high-$\alpha$ and low-$\alpha$ disc and inner and outer halo. Using sophisticated model atmospheres and treatment of NLTE, they derive both 1D NLTE and 3D NLTE abundances for both oxygen and iron across the MW components. Note that although \citetalias{amarsi2019} only classify their stars kinematically using the Toomre Diagram \citep{sandage1987, carney1988} utilizing the three azimuthal velocity components ($\sqrt{U^{2}+W^{2}}$ vs. V), we find good agreement with our more detailed dynamical classifications. 

Prior to assessing the agreement between the GALAH abundances and the \citetalias{amarsi2019} abundances across different components, we corrected our GALAH sample for temperature-dependent abundance trends. Recently, \citet{kos2025} identified these trends in the GALAH DR4 sample by selecting star clusters to act as mono-metallic, chemically homogeneous systems (in the case of open clusters) for which the abundances are expected to be consistent across stellar types. They found clear trends with temperature across abundance space in the clusters but were unable to identify the cause of the trends. Although we selected our stars from GALAH DR3, we identified moderate temperature-dependent abundance trends in our low- and high-$\alpha$ samples, weak trends in our sample of inner halo stars and no significant trend in our accreted halo sample. To correct for these temperature trends, we binned each component in metallicity space into 0.3~dex sized bins to approximate near mono-metallic populations and fit the recovered trends with linear functions to flatten the abundances in each bin across temperature. The de-trended [O/Fe]$_{\mathrm{K20}}$ abundances were carried forward for all further analysis.

Overall, we find good agreement between the mean trends in the GALAH data (after removing the temperature dependence) and the \citetalias{amarsi2019} abundances for each MW structural component. However, after bringing both data sets onto the K20 abundance scale, we still find a 0.25~dex offset in [O/Fe]$_{\mathrm{K20, 1D, NLTE}}$ between all GALAH and \citetalias{amarsi2019} components. We choose only to compare against the 1D results because GALAH DR3 only published 1D NLTE abundances for O and Fe. Given its thorough treatment of abundance uncertainties, we adopt \citetalias{amarsi2019} as the ``ground truth'' and shift the entire GALAH sample \emph{down} by 0.25~dex in [O/Fe]$_{\mathrm{K20}}$. Both the shifted GALAH data and the \citetalias{amarsi2019} data are shown in the left column of Fig.~\ref{fig:galah_amarsi}. The GALAH data are shown as distributions, with the mean trends marked as dashed lines, while the individual abundance measurements are plotted over top as coloured circles. Note the good agreement across components after shifting to the K20 Solar scale and applying a global shift to the GALAH data.

Given the agreement between the mean trends for each component (shown as the dashed line in Fig.~\ref{fig:galah_amarsi}), we interpret the offset as being associated with oxygen only. Unfortunately, we cannot calibrate the offset directly, because there are only two stars in-common between our GALAH sample and \citetalias{amarsi2019}. To test if this offset is related to the fact that our GALAH samples are dominated by giant stars (by design), while the \citetalias{amarsi2019} sample is comprised of dwarf stars (T$_{\mathrm{eff, ave}}\sim5900$~K,  log~$g_{\mathrm{ave}}\sim4.0$), we selected stars within our GALAH components with T$_{\mathrm{eff}}\geq5500$~K and log~$g\geq4$ and re-determined the offset between GALAH and \citet{amarsi2019}. The offset decreased slightly to 0.15~dex, when comparing dwarf stars only, but remained present.

The origin of the offset may be related to the choice by GALAH to determine a single, global zero-point for stars across metallicities. That is, they only investigate abundance accuracy and determine a zero-point offset for Sun-like stars at $\mathrm{[Fe/H]}\sim0$ \citep{galahdr3}, which they then apply to stars across all metallicities.

After comparing the K20-shifted [O/Fe] and [Fe/H] abundances in our GC sample from \citet{carretta2009} with the \citetalias{amarsi2019} high-$\alpha$ disc, in-situ halo and accreted halo components (none of our 16 MW GCs are in the low-$\alpha$ disc), we find a $\sim0.15$~dex offset between \citetalias{amarsi2019} and the GCs in [O/Fe]$_{\mathrm{K20}}$ across the different components. Assuming that first generation stars in GCs (see Sec.~\ref{sec:gcabunds}) in each MW structural component should follow the abundance trends found in field stars in that same component \citep[e.g., as shown in][]{belokurov2024, monty2020}, we again take the \citetalias{amarsi2019} data to be the ground truth shift and the entire GC sample \emph{upwards} in [O/Fe]$_{\mathrm{K20}}$ by +0.15 dex. The disagreement is on the order twice the average uncertainty in the individual [O/Fe] and [Fe/H] abundance measurements ($\sim0.06$ and $\sim0.09$, respectively) reported in \citet{carretta2009}.  

\section{Cosmic Noon Galaxies}
\label{sec:highzcomp}
The study of galaxy enrichment beyond $z\sim0$ has advanced significantly over the last decade, as spectroscopic surveys of $z\sim2-3$ galaxies have increased in size and quality \citep[e.g.,][among others]{steidel2014,kriek2015,mclure2018_vandels}. These surveys have allowed us to learn not only about bulk metallicity but also abundance patterns in the first few Gyr of cosmic time. In recent years, JWST has pushed the frontier of (primarily gas-phase) abundance ratio measurements even further, to $z\sim8$ and beyond \citep[corresponding to the $\lesssim650$~Myr after the Big Bang; e.g.,][]{arellano-cordova2022}.

To assemble a representative superset of abundance measurements for star-forming galaxies in the early Universe, we drew from studies of both individual high-$z$ galaxies and composite spectra as described below. The observable spectra of these galaxies are dominated by light from massive stars and the surrounding star-forming (\ion{H}{II}) regions. At non-ionising rest-UV wavelengths ($\lambda_{\rm rest}\approx1000-2000$~\AA), we directly observe continuum radiation from O and B stars, imprinted with photospheric absorption, stellar wind features, and interstellar absorption lines; in the rest-optical ($\lambda_{\rm rest}\approx3500-7000$~\AA), the ionising radiation from the same stars is reprocessed into hydrogen recombination lines and collisionally-excited forbidden lines of heavier elements such as N, O, Ne, and S. These lines provide detailed information about the physical conditions in the ionised gas in galaxies, including elemental abundances.

It is useful to recall that Fe contributes the majority of the opacity in stellar atmospheres and has the greatest impact on the resulting stellar spectrum, so it is reasonable to assume that $Z_{\ast}\approx Z_{\rm Fe}$. At the same time, the conditions in the gas are closely tied to the abundance of O because oxygen provides some of the most important and efficient cooling pathways, so that $Z_{\rm neb}\approx Z_{\rm O}$. To first order, the enrichment of the stars and gas in \ion{H}{II} regions is the same, owing to the young ages and short lifetimes of the massive stars responsible for both the rest-UV continuum and the rest-optical emission lines. Consequently, it is common to adopt $Z_\ast$ as the level of Fe enrichment and $Z_{\rm neb}$ as the level of O enrichment \emph{for both the massive stars and the gas}. This is the convention that we follow. We also focus primarily on samples where O and Fe were measured for the same galaxies. A summary table of abundances for the Cosmic Noon sample described in this section can be found in Table~\ref{tab:cosmic_noon_abund}.

\subsection{NIRVANDELS}
\label{subsec:nirvandels}

Our principal comparison sample of distant galaxies comes from \citet{cullen2021} and \citet{stanton2024}, who determined O and Fe abundances for a combined sample of $65$ star-forming galaxies at $3.0 < z < 3.8$ (corresponding to lookback times of $11.3-11.9$~Gyr). These galaxies were initially selected from the VANDELS survey\footnote{\url{https://doi.org/10.18727/archive/53}} \citep{mclure2018_vandels, pentericci2018_vandels, garilli2021_vandels}, an ultra-deep optical spectroscopic survey of the CANDELS CDFS and UDS fields undertaken using the VIMOS spectrograph \citep[][$R\sim2500$]{lefevre2003} on ESO’s Very Large Telescope (VLT).
The primary VANDELS targets were typical ``main-sequence'' (i.e., typical SFR for a given $M_{\ast}$) star-forming galaxies for which ultra-deep ($20-80$ hour) rest-frame UV were obtained. Rest-optical follow-up for a subset (dubbed NIRVANDELS) was later obtained with the Keck/MOSFIRE \citep[][$R\sim3500$]{mclean2012} and VLT/KMOS \citep[][$R\sim1985$]{sharples2013} spectrographs, allowing key, abundance-sensitive \ion{H}{ii} region features to be observed. Detailed explanations of the sample selection, observations and data reduction can be found in \citet{cullen2021} and \citet{stanton2024}. 

The Fe abundances were determined via full spectral fitting to the rest-frame FUV spectra using stellar population spectra generated using the high-resolution WM-Basic Starburst99 (S99) models without stellar rotation \citep{leitherer2010_wmbasic, leitherer2014}, assuming a constant SFH over 100 Myr; these models assume $Z_{\odot}=0.014$ and A(Fe)$_{\odot}=7.50$. The forward-modeling approach adopted by NIRVANDELS included a flexible UV dust attenuation prescription, and the posterior distribution was sampled using nested-sampling software \textsc{dynesty} (\citealp{speagle_dynesty}); full details of the spectral fitting approach are given in \citealp{cullen2019}.

The O abundances were estimated from the [O\,{\sc ii}]$\lambda$3727, [Ne\,{\sc iii}]$\lambda$3869, $\rm{H}\beta$ and [O\,{\sc iii}]$\lambda$5007 rest-frame optical emission lines using the \citet{bian2018} strong-line calibrations, which are $T_e$-based diagnostics calibrated using $z\sim0$ galaxies that occupy the same region as $z\sim2-3$ galaxies in the so-called ``BPT diagram'' (which compares log$_{10}$([\ion{N}{ii}]/H$\alpha$) and log$_{10}$([\ion{O}{iii}]/H$\beta$); \citealt{baldwin1981,veilleux1987}). Some of the individual NIRVANDELS galaxies have line ratios outside the range calibrated by \citet{bian2018} and require extrapolation, which may introduce some uncertainty.

\begin{figure}
    \begin{centering}
    \includegraphics[width=\linewidth,trim={0 0.25cm 0 0.2cm},clip]{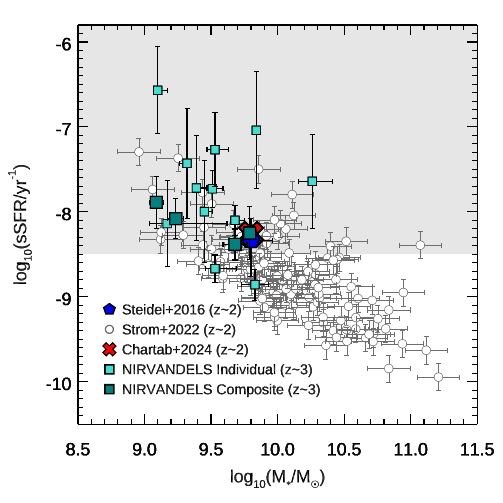}
    \caption{The specific star formation rates (sSFR~$\equiv$~SFR$/M_{\ast}$) and stellar masses (M$_{\ast}$) for the Cosmic Noon samples at $z\sim2$ (at a lookback time of $\sim10.8$~Gyr) and $z\sim3$ (at a lookback time of $\sim11.3$~Gyr). Measurements from composite spectra reported by \citet{steidel2016} and \citet{chartab2024} are shown as the blue pentagon and red cross, respectively; the full KBSS galaxy sample from \citet{strom2022} is shown as the open circles. Individual galaxies and composite spectra from NIRVANDELS \citep{cullen2021,stanton2024} are shown as the small light teal squares and large dark teal squares, respectively. Many of these galaxies have sSFR~$\gtrsim-8.5$ (highlighted by the grey shaded band), which imply formation timescales less than expected minimum Type Ia SNe delay times. It follows that the chemistry of these galaxies would be dominated by CCSNe enrichment \citep{chruslinska2024}.}
    \label{fig:ssfr}
    \end{centering}
\end{figure}

Of the full sample, 12/65 galaxies yielded measurements of both O and Fe abundances, with 38/65 individual determinations of O and 17/65 individual determinations of Fe. To leverage the full sample, four composite spectra were created via stacking, from which robust O and Fe abundances were measured. In our present analysis, we make use of the 12 individual galaxies with both O and Fe abundance measurements, as well as measurements from the four composite spectra. The basic demographic properties of the individual NIRVANDELS galaxies and composite spectra used in the present study are shown in Fig.~\ref{fig:ssfr}, where both sSFR and $M_{\ast}$ are reported assuming a \citet{chabrier2003} stellar initial mass function (IMF). Many of these galaxies have high sSFRs ($\gtrsim-8.5$), consistent with short gas depletion timescales and young ages.

To accurately compare the NIRVANDELS galaxies with the local samples introduced in the previous section, we must place the reported abundances on the \citetalias{kobayashi2020} scale. Fortunately, the electron temperature ($T_e$), or ``direct,'' method of determining O abundances does not assume an intrinsic solar scale but instead uses measurements of the oxygen lines (of various ions) relative to hydrogen recombination lines and the emissivities calculated using $T_e$ and $n_e$ to directly determine O/H. As a result, strong-line diagnostics based on $T_e$ samples already produce absolute A(O) values, so that [O/H]$_{\textrm{K20}}=$~A(O)$-$A(O)$_{\odot,\textrm{K20}}$.

To calculate the absolute Fe abundances from the model comparisons, we use the following equation:
\begin{equation}
\begin{split}
    \textrm{A(Fe)} &= \log_{10}(Z_{\ast}/Z_{\odot,\textrm{S99}})+\textrm{A(Fe)}_{\odot,\textrm{S99}}. 
\end{split}
\end{equation}
To get [Fe/H]$_\textrm{K20}$, we then subtract A(Fe)$_{\odot,\textrm{K20}}$. In this case, because $\textrm{A(Fe)}_{\odot,\textrm{S99}}=\textrm{A(Fe)}_{\odot,\textrm{K20}}$, we can simply adopt [Fe/H]$_\textrm{K20}=\log_{10}(Z_{\ast}/0.014)$.

\subsection {Keck Baryonic Structure Survey}

\citet{steidel2016} inferred O and Fe abundances using composite rest-UV and rest-optical spectra of a sample of 30 star-forming galaxies at $\langle z\rangle=2.40\pm0.11$ (e.g., a lookback time of 10.8~Gyr) that were observed as part of the Keck Baryonic Structure Survey \citep[KBSS;][]{rudie2012,steidel2014}. KBSS comprises 15 fields containing luminous quasars at $z\gtrsim2.7$ and rest-UV color-selected galaxies at $1.5\lesssim z\lesssim3.5$ that were targeted for spectroscopic follow-up using Keck/LRIS \citep[][$R\sim1500$]{oke1995,steidel2004} and Keck/MOSFIRE. The galaxies that were used to construct the composite spectrum are representative of the parent sample, reflected in the location of the \citet{steidel2016} point (blue pentagon) relative to a larger sample of KBSS galaxies (open circles) in Fig.~\ref{fig:ssfr}. Detailed descriptions of the sample selection, observations, and spectroscopic data reduction can be found elsewhere \citep[e.g.,][]{steidel2010,steidel2014,strom2017}, but we review the abundance measurements here.

As for NIRVANDELS, the Fe abundance for the \citet{steidel2016} stack was determined by comparing the FUV composite spectrum with dust-reddened model spectra from both S99 and BPASS \citep[Binary Population and Spectral Synthesis, v2.0;][]{eldridge2016,stanway2016}. Notably, BPASS assumes different solar values from S99, using $Z_{\odot}=0.02$ and A(Fe)$_{\odot}=7.56$. The comparison between the \citet{steidel2016} stack and the stellar population synthesis models spanned a range in stellar metallicity $Z_{\ast}$, IMF slopes, and upper-mass cutoffs for the IMF. Using a $\chi^2$-minimization technique, \citet{steidel2016} found that the observed stellar photospheric absorption and wind lines were best matched by models with stellar metallicities in the range $Z_{\ast}=0.001-0.002$ for both S99 and BPASS. The authors ultimately adopt the BPASSv2.0 models with an upper-mass cutoff of 300~M$_{\odot}$ and an IMF index of $-2.35$. Using the same translation described above, [Fe/H]~$=\log_{10}(0.0015/0.02)+7.56-7.5=-1.06^{+0.12}_{-0.18}.$

The O abundance is determined using the bright emission lines in the rest-optical spectrum by comparing the measured line fluxes (relative to H$\beta$) with predictions from Cloudy \citep[v13.02;][]{ferland2013} photoionisation models. The Cloudy models use the best-fit model stellar spectrum from BPASS as the input ionising radiation field and allow the gas-phase metallicity ($Z_{\rm neb}$) and the ionisation parameter ($U\equiv n_\gamma/n_H$) to vary. These models assumed a plane-parallel geometry and $n_H=300$~cm$^{-3}$, to be consistent with the electron densities\footnote{For ionised gas, $n_H\approx n_e$.} measured for KBSS galaxies \citep{strom2017}. The authors find that the photoionisation model predictions for $Z_{\rm neb}/Z_{\odot}\approx0.5$ and $\log U=-2.8$ are in good agreement with the location of the composite spectrum in multiple line-ratio diagrams. Because Cloudy assumes A(O)$_{\odot}=8.69$, this results in a ``best-fit'' gas-phase A(O)~$=8.38\pm0.10$.

We also use chemical abundance measurements for a sample of individual star-forming galaxies from KBSS \citep{strom2022}, which are shown in Fig.~\ref{fig:ssfr} as open circles. This study used a custom photoionisation model method introduced by \citet{strom2018} to simultaneously infer O, N, and Fe abundances, as well as ionisation parameter $U$, for $\sim200$ individual galaxies. In this method, the full suite of rest-optical emission lines is compared with predicted nebular spectra from Cloudy to determine the most likely combination of $Z_{\rm neb}, Z_{\ast}, U$, and $\log(\rm N/O)$; the input stellar population synthesis models are taken from BPASS, the same as for \citet{steidel2016}. $Z_{\rm neb}$ and $Z_{\ast}$ are converted to O/H and Fe/H as already described. \citet{strom2022} use these measurements to investigate multiple elemental abundance scaling relations with respect to $M_{\ast}$ (i.e., the mass-metallicity relation, or MZR), finding that the O-MZR and Fe-MZR have similar slopes, but that the Fe-MZR is offset toward lower abundances relative to solar. In other words, KBSS galaxies have significantly less Fe than O at all $M_{\ast}$, compared to the Sun.

The two MZRs are expressed as $\textrm{[O/H]}=\alpha_\textrm{O}+\beta_\textrm{O}x$ and $\textrm{[Fe/H]}=\alpha_\textrm{Fe}+\beta_\textrm{Fe}x$, where $x=\log(M_{\ast}/M_{\odot})-10$. Using these equations, we can construct a ``locus'' in the Tinsley-Wallerstein diagram with the following form:
\begin{equation}
\begin{split}
\textrm{[O/Fe]} = \left(\alpha_\textrm{O}-\alpha_\textrm{Fe}\right)+\left(\frac{\beta_\textrm{O}}{\beta_\textrm{Fe}}-1\right)\left(\textrm{[Fe/H]}-\alpha_\textrm{Fe}\right).
\end{split}
\end{equation}
After accounting for the difference between the $Z_{\odot}=0.014$ adopted by \cite{strom2022} and $Z_{\odot,\textrm{BPASS}}=0.02$, and shifting onto the \citetalias{kobayashi2020} scale, the KBSS MZRs have the following parameters:
\begin{equation}
\begin{split}
\alpha_\textrm{O}&= -0.41\pm0.02\\
\beta_\textrm{O} &= 0.14\pm0.05\\
\alpha_\textrm{Fe} &= -0.78\pm0.02\\
\beta_\textrm{Fe} &= 0.17\pm0.05,
\end{split}
\end{equation}
so that $\textrm{[O/Fe]}=(0.37\pm0.03)-(0.18\pm0.08)\left(\textrm{[Fe/H]}+0.78\right)$.

\subsection{Ly\texorpdfstring{$\alpha$}{man Alpha} Tomography IMACS Survey}

\citet{chartab2024} also measure A(Fe) for Cosmic Noon galaxies, applying the method first introduced by \citet{steidel2016} and later employed by NIRVANDELS to a sample of $\sim3000$ star-forming galaxies at $z\sim2.5$ (corresponding to a lookback time of $\sim10.9$~Gyr) observed as part of the Ly$\alpha$ Tomography IMACS Survey (LATIS). As the survey name implies, the rest-UV spectra of the galaxies were observed with Magellan/IMACS \citep[][$R\sim1000$]{dressler2011}. \citet{chartab2024} create a composite spectrum of the entire sample (with the average sSFR and M$_\ast$ shown as the red cross in Fig.~\ref{fig:ssfr}) and use BPASS models to determine the stellar metallicity. They find an average [Fe/H]~$=-0.94\pm0.01$, after accounting for the difference between the $Z_\odot=0.0142$ they assume and $Z_{\odot,\textrm{BPASS}}$ and shifting onto the \citetalias{kobayashi2020} scale. Based on a composite rest-optical spectrum of 17 LATIS galaxies that were previously observed with Keck/MOSFIRE \citep{kriek2015}, they find A(O)~$=8.39\pm0.05$, using the N2 calibration from \citet{bian2018}. 

\begin{table*}
	\centering
	\caption{Summary of the cosmic noon galaxy abundances as absolute abundances ($\mathrm{A(X)} = 12 + \log_{10}\mathrm{(X/H)}$) and placed on the \citealt{kobayashi2020} Solar scale, where $\mathrm{A(Fe)}_{\odot}=7.50$ and $\mathrm{A(O)}_{\odot}=8.76$. Note that the dust depletion correction has been applied to the oxygen abundances listed in columns six and eight. The uncertainty on the correction (0.03~dex) has been included in the quoted uncertainties in these columns, adding the uncertainty in quadrature with the nominal abundance uncertainty. The same uncertainties listed in columns five and six can be applied to the corresponding Solar-scaled abundances. To include more points from KBSS \citep{strom2022}, the following equation yields $\mathrm{[O/Fe]}_{\mathrm{K20}}$ abundances for $\mathrm{[Fe/H]}_\mathrm{K20}$ values in the range of [-0.95, -0.61], $\mathrm{[O/Fe]}_{\mathrm{K20}}=(0.48\pm0.03)-(0.18\pm0.08)\times(\mathrm{[Fe/H]}_\mathrm{K20}+0.78)$.}
	\label{tab:cosmic_noon_abund}
	\begin{tabular}{cccccccc} 
		\hline
		Survey & Galaxy & $z$ & Type &  A(Fe) & A(O) & $\mathrm{[Fe/H]}_{\mathrm{K20}}$ & $\mathrm{[O/Fe]}_{\mathrm{K20}}$\\
		\hline
KBSS        & LM1      & 2.40  & stack        	&  6.44$_{-0.18}^{+0.12}$ 	& 8.48$_{-0.10}^{+0.10}$ & -1.06   & 0.78  \smallskip\\
LATIS       & ...      & 2.50    & stack      	&  6.56$_{-0.01}^{+0.01}$ 	& 8.49$_{-0.06}^{+0.06}$ & -0.94   & 0.67  \smallskip\\
NIRVANDELS  & KVS\_055 & 3.09 & individual 		&  6.83$_{-0.06}^{+0.06}$ 	& 8.73$_{-0.07}^{+0.06}$ & -0.67   & 0.64  \smallskip\\
NIRVANDELS  & KVS\_085 & 3.19 & individual 		&  6.57$_{-0.07}^{+0.09}$ 	& 8.47$_{-0.09}^{+0.08}$ & -0.93   & 0.64 \smallskip\\
NIRVANDELS  & KVS\_204 & 3.47 & individual 		&  6.89$_{-0.06}^{+0.06}$ 	& 8.37$_{-0.10}^{+0.09}$ & -0.61   & 0.21  \smallskip\\
NIRVANDELS  & KVS\_208 & 3.19 & individual 		&  6.52$_{-0.03}^{+0.03}$ 	& 8.42$_{-0.08}^{+0.07}$ & -0.98   & 0.64 \smallskip\\
NIRVANDELS  & KVS\_227 & 3.61 & individual 		&  6.44$_{-0.04}^{+0.04}$ 	& 8.36$_{-0.08}^{+0.08}$ & -1.06   & 0.66 \smallskip\\
NIRVANDELS  & KVS\_248 & 3.08 & individual 		&  6.89$_{-0.06}^{+0.07}$ 	& 8.54$_{-0.07}^{+0.07}$ & -0.61   & 0.39 \smallskip\\
NIRVANDELS  & KVS\_312 & 3.24 & individual 		&  6.67$_{-0.03}^{+0.04}$ 	& 8.38$_{-0.09}^{+0.08}$ & -0.83   & 0.45 \smallskip\\
NIRVANDELS  & KVS\_391 & 3.07 & individual 		&  6.56$_{-0.05}^{+0.04}$ 	& 8.46$_{-0.07}^{+0.07}$ & -0.94   & 0.74 \smallskip\\
NIRVANDELS  &   34777 & 3.40 & individual 		&  6.74$_{-0.05}^{+0.05}$  	& 8.29$_{-0.24}^{+0.24}$ & -0.76   & 0.28  \smallskip\\
NIRVANDELS  &   35212 & 3.40 & individual 		&  7.00$_{-0.09}^{+0.07}$  	& 8.51$_{-0.18}^{+0.17}$ & -0.50   & 0.25  \smallskip\\
NIRVANDELS  &   33644 & 3.20 & individual 		&  6.81$_{-0.08}^{+0.06}$  	& 8.42$_{-0.19}^{+0.18}$ & -0.69   & 0.35  \smallskip\\
NIRVANDELS  &   46857 & 3.35 & individual 		&  6.78$_{-0.04}^{+0.04}$  	& 8.43$_{-0.17}^{+0.18}$ & -0.72   & 0.39  \smallskip\\
NIRVANDELS  &      k1 & 3.50 &  stack 			&  6.55$_{-0.05}^{+0.05}$  	& 8.33$_{-0.07}^{+0.06}$ & -0.95   & 0.53  \smallskip\\
NIRVANDELS  &      k2 & 3.50 &  stack 			&  6.73$_{-0.07}^{+0.07}$  	& 8.47$_{-0.05}^{+0.05}$ & -0.77   & 0.48  \smallskip\\
NIRVANDELS  &      m1 & 3.50 &  stack 			&  6.61$_{-0.06}^{+0.06}$  	& 8.25$_{-0.07}^{+0.07}$ & -0.89   & 0.38  \smallskip\\
NIRVANDELS  &      m2 & 3.50 &  stack 			&  6.78$_{-0.07}^{+0.07}$  	& 8.48$_{-0.07}^{+0.07}$ & -0.72   & 0.44  \smallskip\\
\hline
	\end{tabular}
\end{table*}

\subsection{Towards a Common Abundance Scale \& Systematic Uncertainties II: High-redshift Sample}
\label{sec:highz_systematics}

The resolved stellar abundances used in this study (Sec.~\ref{sec:mwabunds}) all rely on the same fundamental technique, namely equivalent width measurements of stellar absorption features associated with specific atomic transitions. In contrast, the O and Fe abundances of star-forming galaxies, including our Cosmic Noon samples, are determined using a variety of techniques and rely on both measurements of the ionised gas in \ion{H}{ii} regions and the stars illuminating that gas. Differences in the methods used to derive their A(O) and A(Fe) can lead to significant---although not always well quantified---systematic uncertainties \citep[see, e.g.,][]{maiolino2019,chruslinska2024}.

Regardless of the method, we must account for metal depletion onto dust grains when determining total gas-phase abundances. In practice, this is often neglected in extragalactic studies, but given our explicit goal of comparing representative abundance patterns in distant and nearby systems, using methods that probe \emph{different} phases, we cannot ignore the metals locked in dust. This is especially important for A(O), as $\sim20\%$ of oxygen can be contained in dust grains, with the exact amount depending weakly on the total amount of oxygen in the gas \citep[e.g.,][]{peimbert&peimbert2010, pena-guerrero2012}. We account for this effect by adopting the scheme from \citet{peimbert&peimbert2010}, where the depletion of O in dust grains is $0.10\pm0.03$ in ``intermediate O/H'' environments ($7.8<12+\log\textrm{O/H}<8.3$) and $0.11\pm0.03$ in ``high O/H'' environments ($12+\log\textrm{O/H}>8.3$). The result of this correction is to \emph{increase} all [O/H] and [O/Fe] values by $\sim0.1$~dex. In contrast to A(O), A(Fe) is unaffected by dust depletion because it is measured directly from the stellar photospheres and is, therefore, already comparable to A(Fe) measured for the resolved stellar populations. The \citetalias{kobayashi2020}-scaled abundance measurements for our Cosmic Noon sample, now corrected for O depletion onto dust, are shown in the centre and right panels of Fig.~\ref{fig:galah_amarsi}.

Aside from depletion onto dust, there are additional factors that may bias abundance determinations in distant galaxies. For example, the choice of stellar population synthesis model and, e.g., assumptions about binary evolution, stellar rotation and winds, and the IMF can influence the measured A(Fe). In cases where multiple models can be directly compared to the non-ionising UV spectrum \citep[e.g.,][]{steidel2016, cullen2019}, $Z_{\ast}$ appears to be moderately sensitive to differences in the models, with systematic offsets of $\sim0.1-0.2$~dex (compare Fe-MZR in the left and right columns in Fig.~\ref{fig:MZR}). Methods relying on the shape of the ionising radiation as a proxy for $Z_{\ast}$ \citep[e.g.,][]{strom2018,strom2022,sanders2020} show larger differences, as predictions for the ionising UV spectrum vary more widely from model to model. Furthermore, \citet{strom2018} found that the [Fe/H] inferred using their photoionisation model method tended to be higher than the [Fe/H] determined from a non-ionising UV composite spectrum of similar galaxies \citep{steidel2016}. For example, both \citet{strom2018} and \citet{steidel2016} analyse $z\sim2$ galaxies from KBSS and find [O/H]~$\simeq-0.38$ at $\log_{10}(M_{\ast}/M_{\odot})=9.8$ (the average $M_{\ast}$ for the composite spectrum). However, these samples are considerably offset in the centre panels of Fig.~\ref{fig:galah_amarsi}. This is because \citet{steidel2016} report a [Fe/H] that is $\sim0.3$ lower than the Fe-MZR from \citet{strom2018} at the same $M_{\ast}$; the effect is to move the \citet{strom2018} locus to lower [O/Fe] and higher [Fe/H] (down and to the right). Interestingly, this difference is greatly reduced when the photoionisation model method is used to analyse the higher S/N composite spectrum itself, perhaps indicating a need for higher quality observations when using this more indirect method to determine A(Fe).

A separate source of uncertainty in A(O) measurements is the choice of strong-line calibration \citep[e.g.,][]{KewleyEllison08,Kewley19, maiolino2019}. These are used in most studies of faint and/or distant galaxies, because neither metal recombination lines nor the $T_e$-sensitive auroral lines are bright enough to be routinely detected in individual objects. Strong-line diagnostics may be calibrated using $T_e$-based abundances or photoionisation models, but those based on auroral lines typically lead to lower A(O) than model-based methods \emph{for the same objects}. Unfortunately, it is difficult to ascertain ``ground truth,'' because there is also a discrepancy between $T_e$-based abundances (and the strong-line diagnostics based on them) and those inferred from oxygen recombination lines. Both photoionisation model and recombination line-based measurements are often significantly higher (by $\approx0.20-0.24$~dex) than methods anchored using $T_e$. This offset is thought to arise due to temperature fluctuations in the ionised gas \citep[e.g.,][]{Peimbert67,croxall2013,mendez-delgado23}: because less enriched gas is hotter and can more efficiently populate the upper energy levels in metal ions, it will produce more auroral line emission, tending to bias A(O) low. The difference between $T_e$-based abundances and recombination line abundances is a longstanding challenge in studies of nearby nebulae, referred to as the Abundance Discrepancy Factor, or ADF \citep[see][]{tsamis2003,garcia-rojas2007}. However, the exact value of the ADF is subject to vigorous debate and likely differs from \ion{H}{II} region to \ion{H}{II} region, depending on the local ISM conditions \citep[e.g.,][]{mesa-delgado2008,pena-guerrero2012}; indeed, the ADF is found to be negligible in some cases \citep[e.g.,][]{toribiosancipriano2017,chen2023}. Consequently, we opt not to incorporate any systematic shift to account for the ADF in our analysis and instead only apply the dust depletion correction introduced at the beginning of this section.

Finally, most studies of distant galaxies have used \emph{locally} ($z\sim0$) calibrated strong-line diagnostics, which can introduce unknown systematic biases when applied to $z\gtrsim0$ galaxies with significantly different physical conditions. \citet{bian2018} attempted to circumvent this issue by identifying local analogues in line-ratio space, although their calibrations differ significantly from those based on photoionisation models of $z\sim2$ galaxies \citep{strom2018}. JWST promised to mitigate these concerns by enabling new \emph{in situ} calibrations for strong-line diagnostics, based on $T_e$-based oxygen abundances from a representative sample of high-$z$ galaxies, and multiple Cycle 1 proposals were approved to pursue this goal. Early attempts based on supersets of all auroral line measurements to date have already been published \citep[e.g.,][]{laseter2024, sanders2024, scholte2025}, although they do not yet extend to the region of parameter space occupied by most $L^{\ast}$ galaxies at $z\sim2-3$; if we use these calibrations, the A(O) measurements for our Cosmic Noon sample appear shift upward by $\sim0.1$~dex (compare the left and centre columns of Fig.~\ref{fig:MZR}). As the community converges on new, more appropriate high-$z$ strong-line calibrations, it will be important to revisit previously studied samples, including those introduced here.

\section{Comparing Abundance Patterns}
\label{sec:abundpatt}

\begin{figure*}
    \centering
    \includegraphics[width=\linewidth]{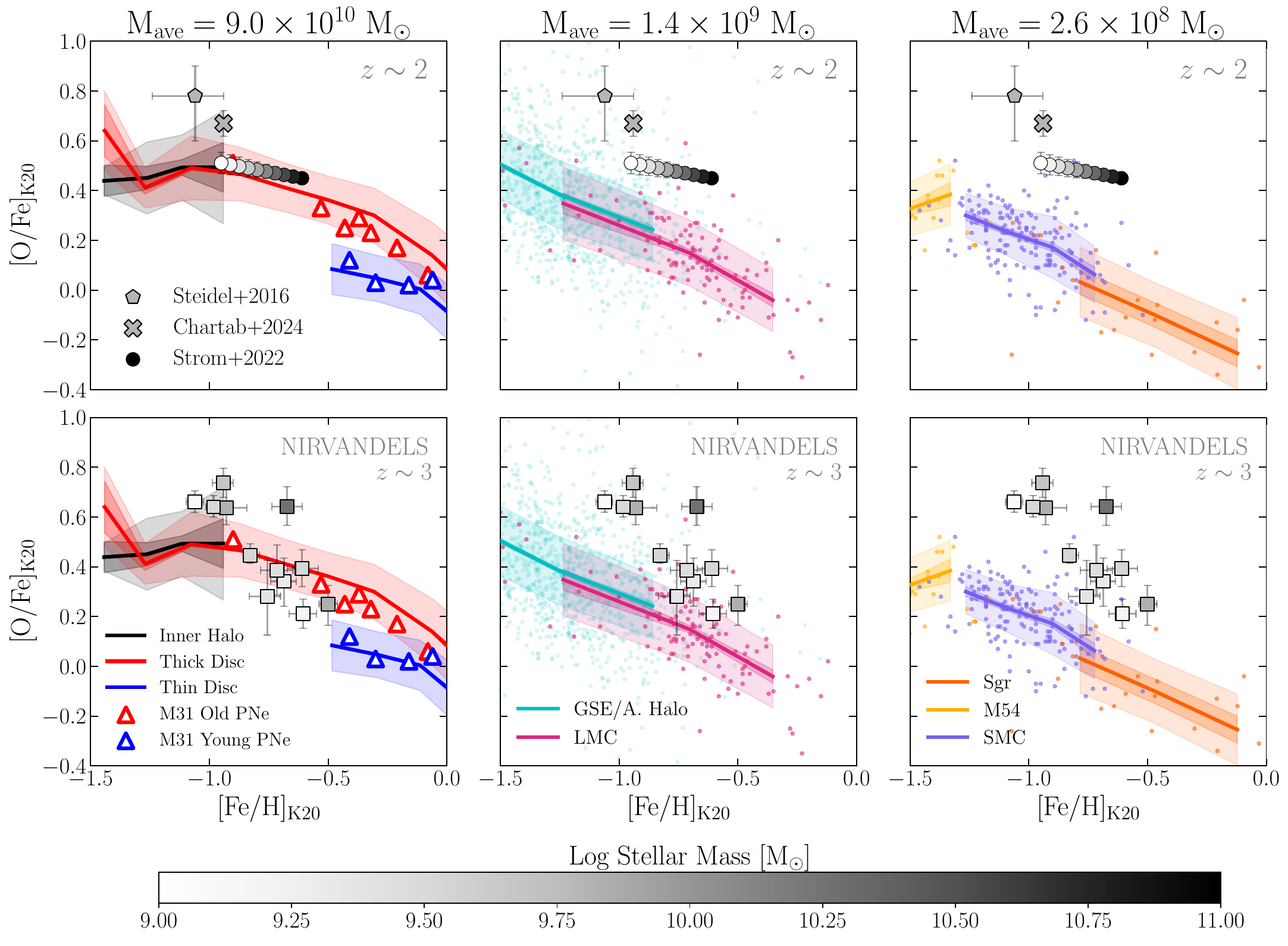}
\caption{The chemical evolution of the MW and four local dwarf galaxies in [O/Fe]$\mathrm{_{K20}}$ vs. [Fe/H]$\mathrm{_{K20}}$ is shown alongside the Cosmic Noon sample (coloured by mass) and data for the two PNe populations in M~31. The $z\sim2$ Cosmic Noon sample is shown in the top row, while the $z\sim3$ sample is shown in the bottom row. The \citet{strom2022} points sample the (oxygen and iron) mass-metallicity relations derived in that study. The title for each panel is the average total $M_{\ast}$ for the local galaxies shown in each (references are given in the text.) \emph{Left,} data for three MW components (from our GALAH sample) represented as the binned mean (bin sizes of 0.2-0.3~dex depending on the component) with the median absolute deviation shown as the shaded region. Medians of O/Ar abundances in the young and old PNe nebulae populations in M~31 in the radial range of 3-14kpc from the study of \citet{arnaboldi2022} and mapped to [O/Fe] and [Fe/H] in \citet{kobayashi2023}, are marked with the blue and red triangles, respectively. \emph{Middle,} the evolution of the accreted halo/GSE (from GALAH, cyan) and the LMC \citep[][pink]{pompeia2008, vanderbarlmc} shown as the binned mean, standard deviation weighted by the number of stars per bin (dark shaded region) and standard deviation (light shaded region). The raw data points are plotted underneath as coloured points. \emph{Right,} same as the central panel except the Sgr dGal (orange) is shown alongside its nuclear star cluster \citep[M~54 (yellow),][]{carretta2010} and the SMC \citep{mucciarelli2023}. \textbf{Note the excellent agreement between the M~31 high- and low-$\alpha$ PNe and the high- and low-$\alpha$ discs in the MW (first column), the appearance of similar [O/Fe]$_{\mathrm{K20}}$ values at fixed metallicities between different galaxies despite distinct star formation histories (columns two and three) and the gradual depletion in [O/Fe]$_{\mathrm{K20}}$ as a function of decreasing stellar mass (all three columns).}}
    \label{fig:lmc_smc_mass}
\end{figure*}

With our local and Cosmic Noon data sets reconciled and placed on the same scale, we are prepared to discuss similarities and differences across the two populations using the Tinsley-Wallerstein diagram. The chemical evolution of our MW components and Local Group galaxies is presented in Fig.~\ref{fig:lmc_smc_mass}. The title of each panel in Fig.~\ref{fig:lmc_smc_mass} is the average total stellar mass (M$_{\ast}$) of the local galaxies included in that panel. We assume a total stellar mass of M$_{\mathrm{MW,\ast}}=5.43\times10^{10}$~M$_{\odot}$ \citep{mcmillan2017} for the MW, M$_{\mathrm{M31,\ast}}=12.5\times10^{10}$~M$_{\odot}$ for M~31 \citep{tamm2012}, M$_{\mathrm{LMC,\ast}}=2.7\times10^{9}$~M$_{\odot}$ for the LMC \citep{shipp2021} and M$_{\mathrm{SMC,\ast}}=3.1\times10^{8}$~M$_{\odot}$ for the SMC \citep{besla2015}. Given both the Sgr and GSE galaxies are either in the process of being disrupted, or are totally disrupted, we assume initial masses of M$_{\mathrm{Sgr,\ast}}=2\times10^{8}$~M$_{\odot}$ for Sgr \citep{vasiliev2021} and M$_{\mathrm{GSE,\ast}}=1.5\times10^{8}$~M$_{\odot}$ \citep{lane2023}, respectively. The Cosmic Noon galaxy compilation is over-plotted in each panel of Fig.~\ref{fig:lmc_smc_mass}, with the $z\sim2$ galaxies shown in the top row and the $z\sim3$ galaxies below. All Cosmic Noon galaxies are shaded by their total stellar mass.

\subsection{\label{sec:cosmicnoon_abund} Cosmic Noon \& Present-Day Galaxies}
For the Cosmic Noon sample, we see clear trends with mass across all three panels in Fig.~\ref{fig:lmc_smc_mass}, consistent with expectations from GCE models. In the $z\sim2$ sample (top row), the intermediate mass \citet{steidel2016} and \citet{chartab2024} points show clear $\alpha$-enhancement, consistent with being massive star-forming galaxies. The KBSS points from \citet{strom2022}, while systematically lower in [O/Fe]$_{\mathrm{K20}}$ for reasons discussed in Sec.~\ref{sec:highz_systematics}, are also significantly $\alpha$-enhanced, even as their overall metallicity increases with increasing mass -- indicative of enhanced SF at comparatively high stellar masses (up to $\sim10^{11}$~M$_\odot$). Similarly, the $z\sim3$ galaxies (bottom row), show a clear gradient of increasing mass with increasing metallicity, which was first noted in \citet{stanton2024} and is again consistent with GCE models. Note that the position of the most massive and chemically evolved NIRVANDELS galaxy, KVS\_055 (the darkest NIRVANDELS data point in Fig.~\ref{fig:lmc_smc_mass}, $\mathrm{M_{\ast}}=1.6\times10^{10}$~M$_{\odot}$) is both at high [O/Fe]$_{\mathrm{K20}}$ and high [Fe/H]$_{\mathrm{K20}}$. 

When we compare the $z\sim2-3$ galaxies and different MW components (shown in the centre and right-most columns in Fig.~\ref{fig:galah_amarsi} and the left-most column of Fig.~\ref{fig:lmc_smc_mass}), we see that they all show good agreement with the high-$\alpha$ disc and $\alpha$-rich inner halo and have significantly higher [O/Fe]$_{\mathrm{K20}}$ at fixed [Fe/H]$_{\mathrm{K20}}$ than the lower mass galaxies shown in the centre and right panels of Fig.~\ref{fig:lmc_smc_mass}. Given the average NIRVANDELS galaxy mass is M$_{\mathrm{ave,*}}=3.35^{+0.87}_{-0.39}\times10^{9}$M$_{\odot}$ at $z\sim3$, a time when the MW was likely significantly smaller, we would expect these galaxies to show a higher level of $\alpha$-enhancement relative to the MW \emph{if} we assume that they have had a constant SFR. If the NIRVANDELS galaxies have experienced a SFH that includes an early strong SF episode, potentially followed by additional bursts, their chemistry may include some contribution from an older population of stars which have been polluted by Type Ia SNe \citep[analogous to the $z\sim1.3-4$ galaxies in $\mathrm{[O/Ar]}$ vs. $\mathrm{[Ar/H]}$,][]{bhattacharya2025, stanton2024b}. We revisit the low-$\alpha$ NIRVANDELS galaxies in upcoming sections, exploring whether different abundance methodologies affect their location in [O/Fe]$_{\mathrm{K20}}$ vs. [Fe/H]$_{\mathrm{K20}}$. 

\subsection{The Milky Way \& M~31}
\label{sec:mwm31}

Comparing the MW to M~31 directly, the left-most panel of Fig.~\ref{fig:lmc_smc_mass} reveals excellent agreement between the MW high- and low-$\alpha$ discs and the M~31 old ($>4.5$~Gyr) and young ($<2.5$~Gyr) PNe, respectively. We interpret this as further support for the existence of high-$\alpha$ and low-$\alpha$ chemical sequences in the inner regions of the disc of M~31 \citep{kobayashi2023}. Note that although both chemical sequences in M~31 are dynamically hot, beyond a galactocentric radius of 14~kpc the older population of PNe display a higher total velocity dispersion than the younger PNe -- suggesting that, as in the MW, the high-$\alpha$ PNe belong to a thick disc population as first noted by \citet{merrett2006} and later confirmed by \citet{dorman2015} and \citet{bhattacharya2019b}. Interestingly, the age ranges provided by the M~31 PNe date both components as being significantly younger than the MW high-$\alpha$ and low-$\alpha$ disc. In the MW, high-$\alpha$ disc ages peak sharply around $\sim11$~Gyr, while the low-$\alpha$ disc displays a broad range of ages, peaking roughly around $\sim5$~Gyr for stars with [$\alpha$/Fe]$<0.2$ using astroseismic ages from \citet{miglio2021}.

While the range of ages is quite small for the PNe associated with the low-$\alpha$ disc of M~31, the range for our own MW low-$\alpha$ disc is substantially broader, signifying extended star formation over $\sim10$~Gyr. 
This opposition is reflected in the high-$\alpha$ PNe population as well, where the broad allowable age range is much larger than the MW high-$\alpha$ disc where \citet{miglio2021} predict SF took place over a period of \emph{only} 1.5~Gyr. Finally, \citet{miglio2021} predict a gap in SF between the two MW discs, which also appears to be the case in M~31 given the two PNe age ranges. 

The origin of the high and low-$\alpha$ sequences in both the MW and M~31 is still the subject of debate. Despite this, the results shown in Fig.~\ref{fig:lmc_smc_mass}, suggest the same outcome (at least chemically) on different timescales. In the case of M~31, results from both cosmological simulations and observations of M~31 GC populations, PNe and RGB stars, suggest a recent, and massive, merger took place $\sim2$~Gyr ago \citep{dsouza2018, hammer2018, mackey2019, dey2023, tsakonas2025}. This is roughly concurrent with the age of the low-$\alpha$ PNe, consistent with PNe kinematics \citep{bhattacharya2023} and driving the two infall GCE model first presented in \citet{arnaboldi2022} and further investigated in \citet{kobayashi2023}. In the MW, the last major merger with the GSE progenitor likely took place after the appearance of the high-$\alpha$, thick disc \citep[explaining the appearance of the ``Splash'' stars, with thick-disc chemistry and heated kinematics,][]{belokurov2020}. Given the range of ages for low-$\alpha$ disc stars in \citet[][see their Figure 13]{miglio2021}, some low-$\alpha$ disc stars may have already formed at the time of the merger. 
While low- and high-$\alpha$ sequences appear in both the MW and M~31 \citep[and in other galaxies, e.g., UGC~10738, ESO~544-27 and IC~1553, where, in every case, dynamically hot high-$\alpha$ populations are found alongside dynamically colder, low-$\alpha$ populations][]{scott2021, somawanshi2024, somawanshi2025}, it is unlikely that the physical cause of the two sequences and any relation to their scale heights is universal.

\subsection{Local Group Dwarf Galaxies \& the GSE}

Moving to the second and third panels in Fig.~\ref{fig:lmc_smc_mass}, we explore similarities between our Local Group dGals. The second panel presents the chemical evolution of our accreted halo sample from GALAH in cyan, with the bulk of the contribution in this metallicity range assumed to be associated with the GSE dGal \citep[e.g.,][]{monty2020, feuillet2020, buder2022, myeong2022}. The LMC is also shown in this plot, combining both the disc and bar populations from \citet{pompeia2008} and \citet{vanderbarlmc}. Note that the population is dominated primarily by the bar population (99/116 stars), which likely represents the most recent episode of star formation in the LMC \citep{vanderbarlmc}. To interpret the similarities and differences between the galaxies in this space, we benefit from the resolved star formation histories of the two systems. 

Beginning with the LMC, both the LMC and SMC have been called ``lazy giants'' due to their relatively delayed SFHs. In the LMC, SF likely peaked only $\sim2$~Gyr ago \citep{nidever2020, hasselquist2021}. Note however that we only see a weak rise in [O/Fe]$_{\mathrm{K20}}$ at high metallicities (related to a burst of SF) seen in the APOGEE data and used to derive the delayed SFH in \citet{hasselquist2021}. As a result of the slow SF, the appearance of the low-$\alpha$ knee in the LMC occurs at very low metallicities, $\mathrm{[Fe/H]}_{\mathrm{K20}}<-2$ \citep{nidever2020}, after which the LMC [O/Fe]$_{\mathrm{K20}}$ values continue to decrease.

The SFH history of GSE likely truncated sometime between 8-10~Gyr ago ($z\sim2$), concurrent with its infall onto the MW \citep{fattahi2019, montalban2021, naidu2021}. However, studies have suggested chemical evidence for a recent starburst in the MW \citep[the component referred to as ``Eos'',][]{myeong2022, ciuca2024, chen2024, matsuno2024, davies2024}, possibly triggered by the infall of GSE. The presence of young stars on GSE-like orbits may support a scenario where star formation was triggered in both the MW and GSE and was more extended than once thought \citep{horta2024}. Regardless of the true SFH of the GSE, be-it early truncated SF, or early formation with a secondary burst triggered by infall onto the MW, it was undeniably more efficent at forming stars at \emph{early} times than the LMC. This is supported by the location of the low-$\alpha$ knee in GSE stars occurring at $\mathrm{[Fe/H]}\sim-1.5$ \citep{myeong2019, monty2020}, 0.5~dex higher than the LMC. 

Finally, in the third panel we show the sequences of Sgr (and its nuclear star cluster, M~54) and the SMC. The SMC also likely had an unusual SFH \citep{nidever2020, hasselquist2021}, with a late stage burst similar to the LMC (and likely related to an interaction between the two). As in the case of GSE, Sgr has a SFH greatly influenced by its interaction with the MW. SFHs derived both photometrically and through chemical evolution modeling agree that the majority of SF in Sgr ceased $\sim5$~Gyr ago, likely related to infall onto the MW \citep{deboer2015,hasselquist2021}. The appearance of secondary star formation episodes (possibly related to continued interaction with the MW) is uncertain. 

In summary, the diversity of SFHs in the Local Group---driving distinct chemical evolution in [O/Fe]$_{\mathrm{K20}}$ vs. [Fe/H]$_{\mathrm{K20}}$, as shown in Fig.~\ref{fig:lmc_smc_mass} and highly affected by their dynamical histories, supports the assertion that galaxies at Cosmic Noon and beyond likely also experienced more complex SFHs beyond simple assumptions of constant SFRs. This point will become increasingly important as JWST observes progressively less massive galaxies at increasing redshifts and we seek to interpret their SFHs.

\section{Chrono-chemical Abundance Trends}
\label{sec:chronotrends}
We now attempt to place our Local and Cosmic Noon samples into the same chrono-chemical context using ages for MW field stars and globular clusters. We focus solely on the NIRVANDELS $z\geq3$ galaxies, based on the assumption that the MW was relatively simple at that time, composed only of the inner halo and early high-$\alpha$ disc, and had not yet merged with GSE. In contrast, based on estimates of the last major merger \citep[10-12~Gyr ago,][]{fattahi2019}, it is likely that the MW evolved significantly at $z\sim2$, so we opt to neglect the $z\sim2$ galaxies in this section. 

\subsection{Look Back Times \& Isochronal Ages}
To convert the NIRVANDELS spectroscopic redshifts to look back times, we use the \texttt{astropy} \texttt{FlatLambdaCDM} \citep{astropy:2022} cosmological model which assumes a flat cosmology, a Hubble constant of 70 km/s/Mpc and a matter density $\Omega_{m}=0.3$ at $z=0$. We assume a total age for the Universe of 13.6~Gyr. Ages for MW stars are taken from the study of \citet{hayden2022} cross-matched with our MW component sub-samples. \citet{hayden2022} determined ages for GALAH DR3 stars using the XGBoost algorithm and a large range of chemical abundances, training the algorithm on isochronal ages for main sequence turn-off stars determined by \citet{sharma2018}. For our sample of GCs we adopt isochronal ages from the study of \citet{vandenberg2013}, noting the consistent use of isochronal ages for both field stars and GCs.

\begin{figure*}
    \centering
    \includegraphics[scale=0.65]{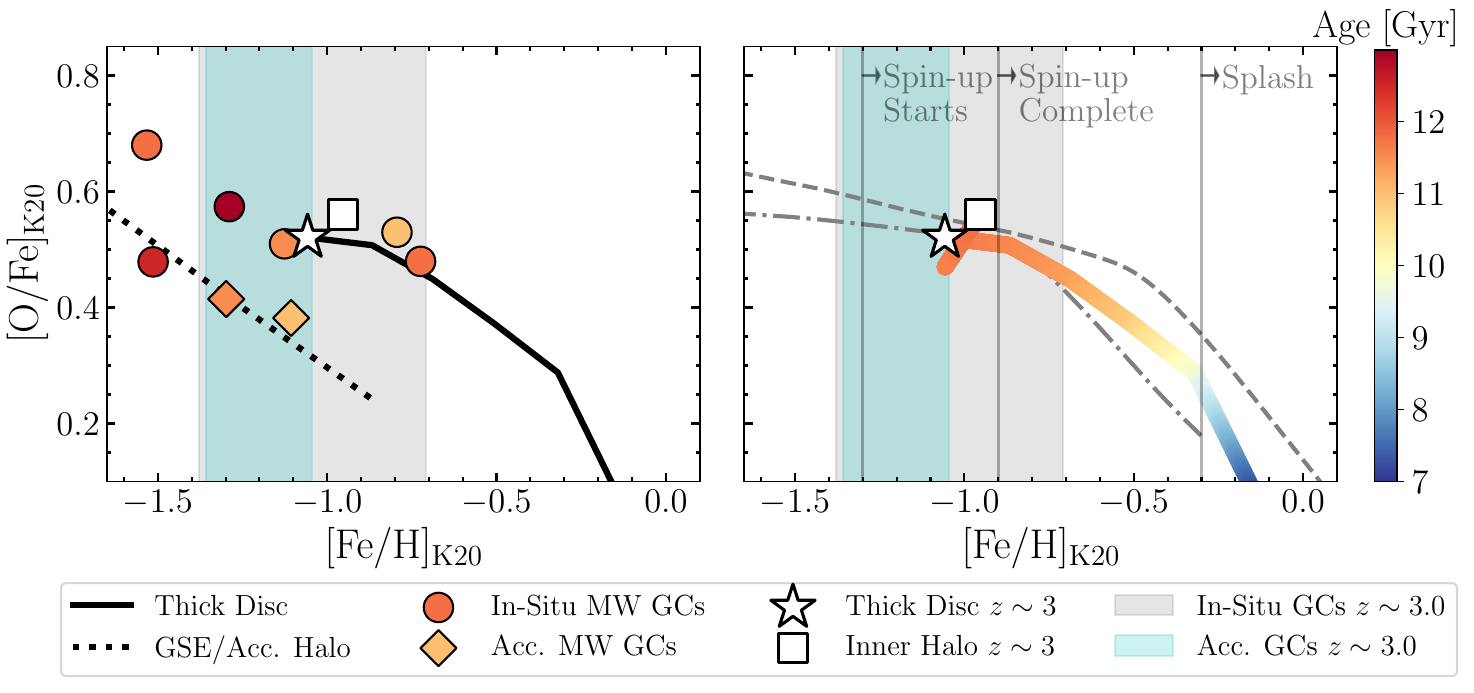}
    \caption{The chrono-chemical evolution of MW inner halo and high-$\alpha$ disc stars (black line in the left panel, rainbow line in the right panel), where the mean age is calculated for 0.2~dex bins in metallicity for both samples to determine the age gradient in the left panel. The location of both samples at $z\sim3$ using the \citet{hayden2022} ages are marked with a square and star, respectively. Ages from \citet{vandenberg2013} are shown for the in-situ GC population (circles), with the location of the clusters shifted by +0.1~dex in [O/Fe]$_{\mathrm{K20}}$, as discussed in Sec.~\ref{sec:mwabunds}. The grey shaded region marks the range in [Fe/H] associated with GC ages in the range of 11.5-12~Gyr (close to $z\sim3$). GCs associated with the GSE merger by \citet{myeong2019} and \citet{massari2019} are marked with diamonds. The cyan shaded region marks the range in [Fe/H] associated with GSE GC ages in the range of 11-11.5~Gyr (the only available ages). GCE models from K20 for the high-$\alpha$ disc (filled line) and bulge (dashed line) are also included. The mean evolution of the GSE sample in GALAH is shown with the dotted line. The estimated location of the beginning and end of the emergence of the high-$\alpha$ disc, or ``spin-up'' from \citet{Belokurov+2022} in metallicity are marked in the left column as vertical lines. The location of the completion of the ``Splash'' heating of the high-$\alpha$ disc in metallicity is also marked via a vertical line to give context to the MW's evolution. \textbf{Note the agreement between the location for the MW at $z\sim3$ in [O/Fe]$_\mathrm{K20}$ and [Fe/H]$_\mathrm{K20}$ predicted by the field stars (square and star) and the population of in-situ GCs (black shaded region) and that these predictions place the MW just prior to the completion of the formation of the high-$\alpha$ (thick) disc at $z\sim3$.}}
    \label{fig:ages}
\end{figure*}

Field star and GC ages are presented in Fig.~\ref{fig:ages}. The mean \citet{hayden2022} ages for both in-situ halo and high-$\alpha$ disc stars, using 0.2~dex-sized bins in [Fe/H]$_{\mathrm{K20}}$ are shown alongside the mean evolution of [O/Fe]$_{\mathrm{K20}}$ vs. [Fe/H]$_{\mathrm{K20}}$. Because \citet{hayden2022} do not derive ages for stars below $\mathrm{[Fe/H]}=-1$, the oldest mean age in our high-$\alpha$ disc sample is coincidentally the average upper limit on the NIRVANDELS ages (11.7~Gyr). The location of the high-$\alpha$ disc (in-situ halo) in [O/Fe]$_{\mathrm{K20}}$ vs. [Fe/H]$_{\mathrm{K20}}$ at 11.7~Gyr is marked with a white star (white square). GCE models for the high-$\alpha$ disc and bulge from K20 are also marked in Fig.~\ref{fig:ages} as the filled line and dashed line respectively. When our high-$\alpha$ disc sample transitions to younger ages, showing stronger disagreement with the GCE model, we are likely dominated by contamination from younger, low-$\alpha$ disc stars which have made it into our high-$\alpha$ disc sample (around $\mathrm{[Fe/H]}_{\mathrm{K20}}\sim-0.5$). 

\subsection{The Milky Way \& GSE at $z\sim3$}
Using the field star and GC ages, we now attempt to project the MW and GSE back in time to $z\sim3$ (11.7~Gyr look back time) to the era of the NIRVANDELS galaxies. Beginning with the in-situ GC ages, selecting only GCs with an isochronal ages between 11-12~Gyr, we place the MW somewhere in the range of $-1.3\leq\mathrm{[Fe/H]}_{\mathrm{K20}}\leq-0.7$ at $z\sim3$. This range is marked with the shaded black region in Fig.~\ref{fig:ages} and reflects the average metallicity and standard deviation of the subset of in-situ GCs. Field star ages from \citet{hayden2022} support this placement, with an average binned age of 11.69~Gyr in the high-$\alpha$ disc and inner halo occurring at $\mathrm{[Fe/H]}_{\mathrm{K20}}=-0.95$ and $\mathrm{[Fe/H]}_{\mathrm{K20}}=-1.06$, respectively (marked as the white star and white square in Fig. ~\ref{fig:ages}, respectively). 

In the case of GSE, only two GSE-tagged GCs in our sample have ages in the range of $z\sim3$ (NGC~2808 and NGC~5904) in our combined \citet{carretta2009}-\citet{vandenberg2013} catalogue. These two GCs have isochronal ages of 11.0~Gyr and 11.5~Gyr, respectively, making the placement of the GSE point in [O/Fe]$_{\mathrm{K20}}$ ([Fe/H]$_{\mathrm{K20}}$) a lower (upper) limit. The range of metallicity spanned by these two GCs is $-1.4\leq\mathrm{[Fe/H]_{\mathrm{K20}}}\leq-1.0$ marked with the cyan shaded region. 

\begin{figure*}
    \centering
    \includegraphics[scale=0.5]{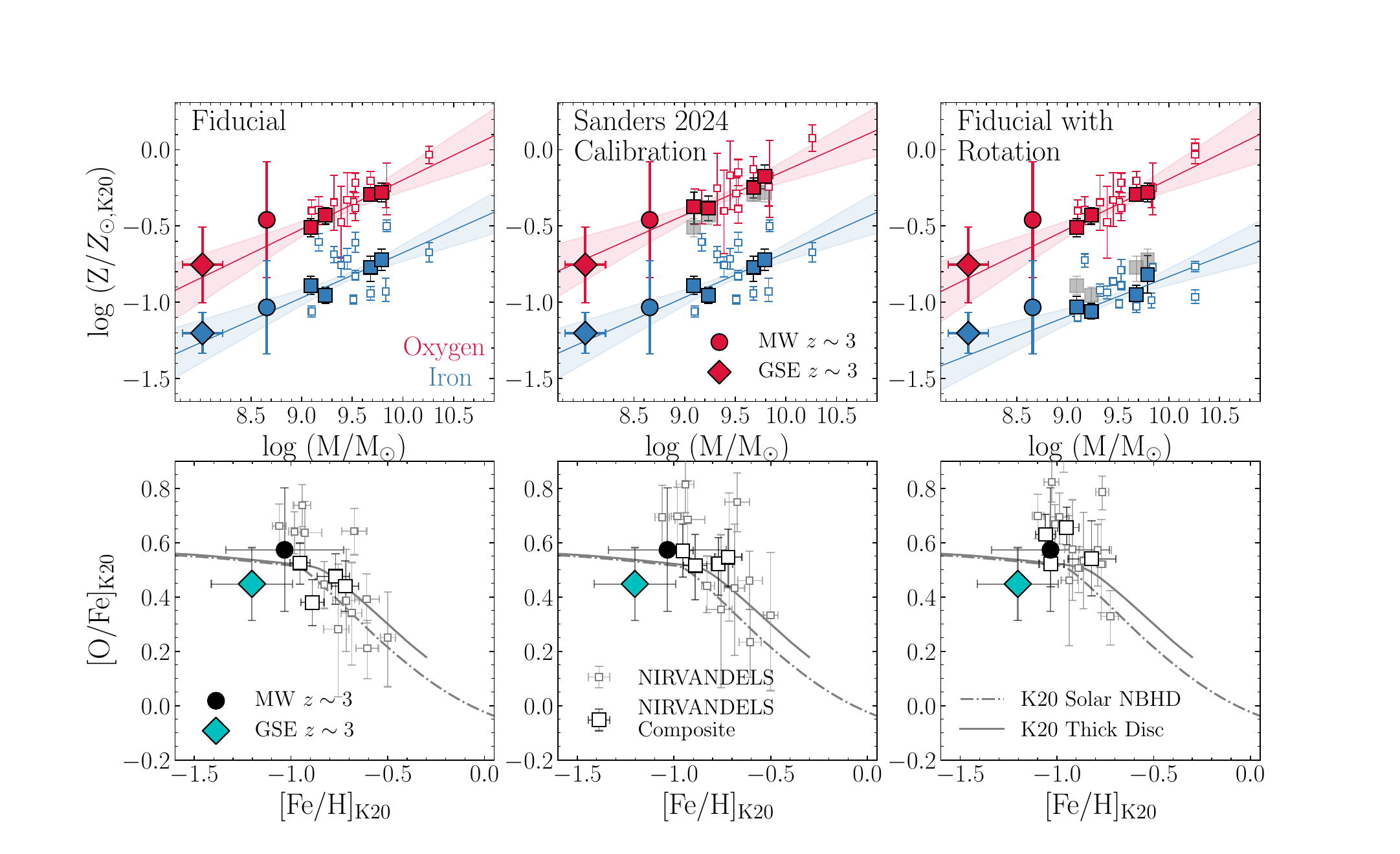}
    \caption{Results of projecting the MW (primarily the inner halo or ``Aurora'') and GSE to $z\sim3$. \emph{Top,} metallicity-mass relation (MZR) using $Z_{\mathrm{gas}}$ (oxygen abundance, red) and $Z_{\mathrm{*}}$ (iron abundance, blue) for the NIRVANDELS sample of individual galaxies (small points) and composites (larger points). The MW and GSE are marked with the circle and diamond, respectively. \emph{Bottom,} Individual and composite NIRVANDELS galaxies in [O/Fe] vs. [Fe/H] alongside the projected location of the MW (black circle) and GSE (teal diamond). The three columns reflect different methodologies to extract the galaxy abundances. In the first and last columns, ${\rm Z_{gas}}$ is determined using the \citet{bian2018} strong-line calibration scheme, while in the middle column ${\rm Z_{gas}}$ is determined using the calibration scheme of \citet{sanders2024}. While all three columns derive ${\rm Z_*}$ using S99 models, the left and middle column use S99 models with no rotation, whilst the right column uses models including rotation. The location of the NIRVANDELS stacks in O and Fe determined using the fiducial calibration scheme (left-most column) are included in the middle and right-most columns (as grey points) to highlight the impact of different calibration schemes. \textbf{Note the remarkable consistency of the MW and GSE at $z\sim3$ with predictions for the evolution of the gas (oxygen)- and stellar (iron) MZR from NIRVANDELS (top row) and the sensitivity of the NIRVANDELS abundances to the choice of calibration scheme (the shift in the appearance of a decreasing trend in [O/Fe]$_{\mathrm{K20}}$ vs [Fe/H]$_{\mathrm{K20}}$ to the appearance of a plateau, across panels in the bottom row).}}
    \label{fig:MZR}
\end{figure*}

To estimate the location of the MW in [Fe/H]$_{\mathrm{K20}}$ and [O/Fe]$_{\mathrm{K20}}$ at $z\sim3$, we average the [Fe/H]$_{\mathrm{K20}}$ and [O/Fe]$_{\mathrm{K20}}$ values from the in-situ clusters with ages between 11-12~Gyr and do the same with the two GSE-tagged GCs to estimate a location for GSE. Our final estimated values are $(\mathrm{[Fe/H]_{K20}, [O/Fe]_{K20}})=(-1.04\pm0.33, 0.55\pm0.18)$ for the MW and $(\mathrm{[Fe/H]_{K20}, [O/Fe]_{K20}})=(-1.20\pm0.13, 0.45\pm0.16)$ for the GSE at $z\sim3$, where the uncertainty is the standard deviation for the set of GCs added in quadrature with the average individual GC measurement uncertainty and the shift applied to bring the GCs into agreement with \citetalias{amarsi2019} on the K20 scale (+0.15~dex, see Sec.~\ref{sec:gcabunds}). 

Given the difficulty in deriving ages for metal-poor stars \citep[$\mathrm{[Fe/H]}<-1$, see ][for a discussion of the complications in deriving astroseismic ages at the star-by-star level]{alancastro2022}, we have limited data sets to compare against for our placement of GSE and the MW in [Fe/H]$_{\mathrm{K20}}$, [O/Fe]$_{\mathrm{K20}}$ at $z\sim3$. We can compare the location of our GSE data points against the recent results of \citet{horta2024} who select likely GSE members from the catalogue of MSTO-isochronal ages of \citet{xiang2022}. \citet{horta2024} find an average age of $11.6^{+1.14}_{-1.15}$~Gyr for their GSE sample across a metallicity range of $-2\leq\mathrm{[Fe/H]}\leq-1$. This range reinforces the likelihood that our placement of GSE at [Fe/H]$_{\mathrm{K20}}=-1.2$ at $z\sim3$ represents an upper limit on the location of GSE (it was likely less evolved than this at that time). 

Another recent study examining the ages of accreted MW stars using both astroseismology and isochrones was performed by \citet{debrito2024}. Examining their accreted stars with $\mathrm{[Fe/H]}<-1$, and neglecting a single outlier with an astroseismic age of 6.5~Gyr, they find an average astroseismic age of 11.4~Gyr at $\mathrm{[Fe/H]}=-1.24$ - in good agreement with our placement based on accreted GCs, but again suggesting that we are overestimating the location of GSE in [Fe/H]$_{\mathrm{K20}}$ at $z\sim3$.

Comparing our placement of the MW in [Fe/H]$_{\mathrm{K20}}$, [O/Fe]$_{\mathrm{K20}}$ at $z\sim3$ with literature results, we find good agreement with the study of \citet{queiroz2023} who derive MSTO-isochronal ages for the high-$\alpha$ disc using their \texttt{StarHorse} code and multiple surveys (see their Figure 11, where 11~Gyr old stars show a mean metallicity of $\mathrm{[Fe/H]}\approx-1$ in the LAMOST DR7, APOGEE DR17 and GALAH DR3 surveys). The results of \citet{silva2018} using the APOKASC catalogue \citep{pinsonneault2014} to derive ages for the high-$\alpha$ and low-$\alpha$ discs, also qualitatively agrees with our placement of the MW in [Fe/H]$_{\mathrm{K20}}$, [O/Fe]$_{\mathrm{K20}}$, though they do not age-date stars with $\mathrm{[Fe/H]}<-1$ as their sample of high-$\alpha$ disc stars truncates around $\mathrm{[Fe/H]}=-0.9$ (on the APOGEE scale, who in-turn adopt the abundances of \citet{grevesse2007} where $\mathrm{A(Fe)}=7.45$, 0.05~dex lower than the K20 scale). In the range of $-0.9\leq\mathrm{[Fe/H]}\leq+0.4$, they find an average astroseismic age of 11~Gyr for the high-$\alpha$ disc, with ages generally increasing with decreasing metallicity (see their Figure 10). 

\subsection{The $z\sim3$ Mass-Metallicity Relation}
Using our estimate of the position of the MW and GSE in [O/Fe]$_{\mathrm{K20}}$, [Fe/H]$_{\mathrm{K20}}$ at $z\sim3$, we aim to place the two galaxies on the mass-metallicity relation (or MZR) for the NIRVANDELS galaxies recovered in \citet{stanton2024}. The MZR was first observed in local galaxies \citep[e.g.,][; see the final reference for a more complete review of the subject]{mcclure68, sandage72, maiolino2019}. Very broadly, the physical cause of the relationship between increasing galactic mass and increasing metallicity is likely due to the shallower potential of low-mass galaxies, which allows SN-driven winds to more efficiently eject their metal-enriched gas \citep[see ][for a recent and more detailed model capturing the many physical mechanisms driving the MZR]{sharda2021b, sharda2021a}. 

The MZR appears to apply to both both quiescent and star-forming galaxies, with different slopes. Its existence has been confirmed for the gas-phase ($Z_{\mathrm{gas}}$, [O/H] and [N/H]) metallicity out to $z\sim3.5$ \citep[e.g.,][]{steidel2014, onodera2016, sanders2021, cullen2021,strom2022, stanton2024}, with JWST surveys providing initial measurements out to $z\sim10$ \citep[e.g.,][]{Nakajima23,Chemerynska24,curti2024}.
The stellar ($Z_{*}$, [Fe/H]) metallicity MZR has been measured out to $z < 5$ \citep[e.g.,][]{cullen2019, kashino2022, strom2022, chartab2024, stanton2024}. The common slopes but differing normalizations for ${\rm Z_{gas}}$ and ${\rm Z_*}$ MZRs reflect the different nucleosynthetic channels responsible for forming oxygen and iron, providing further evidence for $\alpha$-enhancement in high-$z$ systems \citep{cullen2021, strom2022, stanton2024}.

To assign an estimate for the mass of the MW and GSE at $z\sim3$, we utilize results from the recent study of \citet{kurbatov2024}. They determine the density profiles for two ancient MW components, ``Aurora'' \citep[what we refer to as the ``inner halo'',][]{Belokurov+2022} and GSE. In the case of GSE, we adopt the results from \citet{kurbatov2024} assuming the exponentially truncated single power-law density profile for the spatial distribution of GSE from \citet{lane2023}. \citet{kurbatov2024} provide estimates of the total stellar mass for each component as a function of metallicity (in the range of $-3\leq\mathrm{[Fe/H]}\leq-1$ using metallicities for Gaia RGB stars from \citealt{andrae2023}). Using our estimates of the location of both the MW and GSE at $z\sim3$ in [Fe/H], we adopt a mass of $4.52\times10^{8}$~M$_{\odot}$  for Aurora (integrating up until $\mathrm{[Fe/H]}=-1$) and $1.05\pm0.24\times10^{8}$~M$_{\odot}$ for GSE (averaging the total integrated mass between $-3\leq\mathrm{[Fe/H]}\leq-1.3$ and $-3\leq\mathrm{[Fe/H]}\leq-1.0$). Note that \citet{kurbatov2024} does not quote uncertainties on their mass estimates.

Because the GCs we use to place GSE in ([O/Fe]$_{\mathrm{K20}}$, [Fe/H]$_{\mathrm{K20}}$) have an average age (11.25~Gyr) less than our desired lookback time (11.7~Gyr), our positioning of GSE is likely an overestimate of how evolved the galaxy was at $z\sim3$. To explore by how much we have have overestimated the mass for the GSE, we turn to two studies which predict the SFHs of GSE using GCE modeling. Looking first at the study of \citet{sandersj2021}, they derive a SFH for GSE using data from APOGEE DR16 \citep{apogeedr16data}, GALAH DR3 and \citet{nissen2010} and a bespoke GCE model. In the range of 11.7-11.25~Gyr, \citet{sandersj2021} place GSE at the peak of its SF. Because of this, their results indicate GSE is approximately 2-3 times smaller at $z\sim3$ than our estimate. The other study we compare to is \citet{hasselquist2021}, who derive derive SFH for GSE using APOGEE DR16 data and two GCE models. In the SFH derived by \citet{hasselquist2021} using the \texttt{flexCE} \citep{flexce} GCE model, between 11.7-11.25~Gyr GSE is only moderately SF, with the peak occurring later at $\sim9$~Gyr \citep[consistent with the second age peak seen in ][]{horta2024}. Based on this, GSE was only 1.1 times more massive at 11.25~Gyr than it was at $z\sim3$.  

As shown in Fig.~\ref{fig:ages} using vertical lines, the appearance of coherent rotation and the emergence of the high-$\alpha$ disc (``spin-up'') begin around $\mathrm{[Fe/H]=-1.3}$ and is complete by $\mathrm{[Fe/H]>-0.9}$ \citep[determined using field star kinematics and chemical abundances, ][]{Belokurov+2022}. Note that the ``spin-up'' phase may have begun even earlier than $\mathrm{[Fe/H]=-1.3}$ based on recent results from \citet{viswanathan2024}. Regardless, our estimate of the $z\sim3$ MW at $\mathrm{[Fe/H]_{K20}}=-0.98$ based on field star and GC ages places the Galaxy slightly before the end of the primary disc-building phase -- meaning that some high-$\alpha$ disc is present in our inner halo, or Aurora, sample. However, because \citet{kurbatov2024} do not define their components in velocity space, contributions from the in-situ high-$\alpha$ disc up until $\mathrm{[Fe/H]}<-1$ are accounted for in their estimate, and our adopted total mass of the inner halo. We do not consider the contribution of the MW bulge in our estimate of the in-situ population of the MW at $z\sim3$ because stellar age estimates are extremely limited for bulge stars, and the metallicity range of bulge stars with reliable estimates of ages greater than 11~Gyr spans 1.5~dex \citep{bensby2017, joyce2023} making it difficult to pinpoint a single location in the MZR. 

We present the oxygen- and iron-MZR derived from the NIRVANDELS galaxies in the top row of Fig.~\ref{fig:MZR} alongside estimates for the location of MW and GSE at $z\sim3$. The best-fit trends from \citet{stanton2024} determined from the NIRVANDELS composite spectra (large squares) are shown for both the oxygen- (red) and iron-MZR (blue). Our estimated positions for the MW and GSE on the MZR are marked with a circle and diamond, respectively. 
Remarkably, the stellar-based oxygen abundances and iron metallicities of both the MW and GSE seem to follow the low mass extrapolation of the corresponding relations obtained by \citet{stanton2024} across the three different modeling techniques (shown across the three columns). Recent results for the $z=3$ MZR from the \textsc{thesan-zoom} \citep{kannan2025} simulations also show very good agreement with the NIRVANDELS MZR and our MW and GSE data points \citep{McClymont2025}. The oxygen- and iron-MZRs in the first column are calculated using the using the fiducial strong-line calibrated scheme and stellar models described in Sec.~\ref{subsec:nirvandels}, the middle column is calculated using the same stellar models but applies the JWST-based calibration scheme of \citet{sanders2024}, while the right-most column uses the fiducial oxygen abundances but incorporates S99 models with rotation. 

The effects of these different approaches were discussed in Sec.~\ref{sec:highz_systematics}, and the bottom row of Fig.~\ref{fig:MZR} illustrates the impact on the location of the NIRVANDELS data points in [O/Fe]$_{\mathrm{K20}}$ vs. [Fe/H]$_{\mathrm{K20}}$. Adopting the JWST-based strong-line diagnostics from \citet{sanders2024} increases the gas-phase O/H by $\sim0.1$~dex (center panels), shifting the $z\sim3$ galaxies above the K20 relation, MW and GSE, in line with expectations of the NIRVANDELS galaxies being more greatly oxygen-enhanced than the MW at similar epochs. Including stellar rotation has the same overall effect by decreasing the stellar Fe/H by $\sim 0.1$ dex (right panels). We note that stellar population models that take into account the presence of binary stars \citep[BPASS,][]{stanway2016,eldridge2016,StanwayEldridge18} have a similar systematic effect on the derived Fe abundances as the use of single-star models with rotation (i.e., it typically leads to lower Fe abundances than those derived using S99 models without rotation, thereby shifting high-$z$ galaxies up and to the left in the [O/Fe] - [Fe/H] diagram, e.g., \citealt{chisholm19,cullen2019,cullen2021}).

The position of the MW and GSE on the star-forming NIRVANDELS MZR suggests that both galaxies were actively star-forming at $z\sim3$. Given the metallicity distributions of both galaxies, they were likely star forming at even earlier times. It is possible that they were among the low-mass, star-forming galaxies which could have reionised the Universe. In their recent study examining 677 galaxies across $z=4-7$ in the mass range of log~(M$_{*}$/M$_{\odot}$)=7-10 (less massive than both the MW and GSE at $z\sim3$), \citet{simmonds2024} find an increasing trend of ionising photon production efficiency with decreasing galaxy UV brightness and stellar mass \citep[see also][]{begley2025}. They conclude that, if their sample is representative of faint, low mass, bursty star-forming galaxies, these galaxies are efficient enough to reionise the Universe. Given that \citet{simmonds2024} were able to observe $10^{7}$~M$_{\odot}$ galaxies at $z>4$ and given our estimated masses for GSE and the MW ($>10^{7}$~M$_{\odot}$ at $z\sim3$), it is possible that JWST is currently observing GSE and MW analogues in the early Universe, an exciting prospect for the placement of our Local Universe in the broader context of global galaxy evolution.

\section{Conclusions}

With the shared goal of understanding the chemical evolution of all galaxies including our own, we founded the ChemZz collaboration to couple results from extragalactic and Galactic chemical abundance studies. In this first study, we placed oxygen and iron measurements made from evolved stars in the Milky Way, Local Group galaxies, and globular clusters onto the same scale as those made in star forming regions in unresolved galaxies at Cosmic Noon ($z\sim2-3$). To do this, we reviewed the assumptions relevant to the methods in both samples and the systematics affecting each. 

Abundances for the Milky Way (MW)  were taken from the third data release of the GALAH survey \citep{galahdr3}, splitting the MW into four chemodynamic components, namely the high-$\alpha$ (thick) and low-$\alpha$ (thin) discs and inner and outer halo, the latter we assume to be dominated by debris from the Gaia-Sausage-Enceladus merger \citep{belokurov2018, helmi2018} using various chemical and kinematic cuts. Abundances from the O-normal, first population, stars in MW globular clusters (GCs) were also included, taken from the study of \citet{carretta2009}. After bringing the MW and GC data onto the K20 scale and considering their kinematic associations, we applied small shifts to bring them into agreement with the study of \citet{amarsi2019} on the \citet[][K20]{kobayashi2020} scale, which we adopt as our ground truth abundances. We also compiled abundance measurements for evolved stars in the LMC \citep{pompeia2008, vanderbarlmc}, SMC \citep{mucciarelli2023} and Sgr \citep{carretta2010} dwarf galaxies and measurements derived from planetary nebulae in M~31 \citep{bhattacharya2022}. 

Extragalactic O and Fe abundances were collated from three main surveys, the Keck Baryonic Structure Survey \citep[KBSS][]{rudie2012, steidel2014}, the Ly$\alpha$ Tomographic IMACS Survey \citep[LATIS][]{chartab2024} and the NIRVANDELS survey \citep{cullen2021, stanton2024}. In the case of LATIS, we adopt the stacked measurement (derived from composite spectra created by combining all individual galaxies in the survey) only, in KBSS we include both the stacked measurement from \citet{steidel2016} and representative points from the mass-metallicity relations derived by \citet{strom2022}, and in NIRVANDELS we include both the stacked and individual galaxy abundances. All abundances were corrected for depletion onto dust and brought onto the K20 scale.

After bringing all of our data sets to a common scale, we compare resolved stellar abundance measurements of oxygen and iron in Local Group galaxies and star clusters, with gas-phase and stellar metallicities in galaxies at Cosmic Noon via the Tinsley-Wallerstein diagram \citep{wallerstein1962, tinsley1979}. \smallskip

\noindent\textbf{The main takeaways from this study are as follows:}

 \begin{itemize}
     \item We find that stacked abundances from the LATIS \citep{chartab2024} and KBSS surveys \citep{steidel2016, strom2022} show clear O-enhancement in agreement with the MW high-$\alpha$ disc and inner halo, while individual massive, star-forming galaxies from NIRVANDELS \citep{cullen2021, stanton2024} show increased diversity of [O/Fe]$_{\mathrm{K20}}$ abundances, agreeing with both the high-$\alpha$ and low-$\alpha$ discs (left panels of Fig.~\ref{fig:lmc_smc_mass}). 
     \smallskip 
     \item When placed onto a common abundance scale, we find excellent agreement in [O/Fe]$_{\mathrm{K20}}$ vs. [Fe/H]$_{\mathrm{K20}}$ between the MW high-$\alpha$ (thick) and low-$\alpha$ (thin) discs and old $\alpha$-rich, and young $\alpha$-poor, planetary nebulae in M~31, supporting the existence of an $\alpha$-bimodality in the inner regions of the disc of M~31 (left panels of Fig.~\ref{fig:lmc_smc_mass}). \smallskip 
     \item Using ages for globular clusters associated with the in-situ MW and accreted halo (assumed to dominated by the Gaia-Sausage-Enceladus, GSE) we estimate the location of the MW and GSE in [O/Fe]$_{\mathrm{K20}}$ vs. [Fe/H]$_{\mathrm{K20}}$ at $z\sim3$, or a look back time of 11.7~Gyr (Fig.~\ref{fig:ages}). Our final estimated values are $(\mathrm{[Fe/H]_{K20}, [O/Fe]_{K20}})=(-1.04\pm0.33, 0.55\pm0.18)$ for the MW and $(\mathrm{[Fe/H]_{K20}, [O/Fe]_{K20}})=(-1.20\pm0.13, 0.45\pm0.16)$ for the GSE.\smallskip 
     \item After re-scaling, and independent of the calibration scheme, some NIRVANDELS galaxies appear more [O/Fe]$_{\mathrm{K20}}$-poor than the less massive MW at $z\sim3$ (bottom panels of Fig.~\ref{fig:MZR}), implying the existence of stars that were formed after Type Ia SNe had contributed a significant amount of Fe. This likely reflects a diversity of star formation histories that may include already declining star formation or an early episode of star formation  followed by additional bursts \citep[e.g., as in the bursty models of][]{kobayashi2024}. \smallskip 
     \item Combining our estimate of the location of the MW and GSE in [O/Fe]$_{\mathrm{K20}}$ and [Fe/H]$_{\mathrm{K20}}$ at $z\sim3$ with estimates of their masses at this time from \citet{kurbatov2024}, we plot the two on the O-, Fe-mass-metallicity relation of star-forming NIRVANDELS galaxies. We find excellent agreement between the location of the two galaxies and the predictions for the low-mass end of both the NIRVANDELS O- and Fe-mass-metallicity relations, independent of the abundance calibration scheme (top panels of Fig.~\ref{fig:MZR}). 
 \end{itemize}

In the future, we will investigate the abundance of carbon and nitrogen across the Local Group and at Cosmic Noon and include measurements of the kinematics of our Cosmic Noon sample to understand the relationship between chemical evolution and the emergence of galactic discs. Additionally, a larger set of homogeneous globular cluster isochronal ages from projects like CARMA \citep{massari2023} will greatly aid with attempts to project both the MW and GSE further back in time in abundance space. Upcoming large scale surveys measuring homogeneous NLTE-corrected oxygen abundances for millions of stars throughout the MW, like WEAVE \citep{weave} and 4MOST \citep{4most}, will greatly aid studies like ours. Further studies like ours and that of \citet{ji2025} and \citet{chruslinska2024}, working to compare resolved stellar abundances directly to stellar- and gas-phase abundances and metallicities across redshifts will continue to test the compatibility of these distinct abundance derivation techniques. In parallel, harnessing techniques like those of \citet{lian2023}, which translate resolved stellar abundances in the MW to integrated light measurements will allow for more direct comparisons between Local Group galaxies and extragalactic abundances. Moreover, integral-field spectrographs on the ELTs will allow us to spatially resolve low-mass galaxies at Cosmic Noon \citep{newman2019}, enabling novel connections between distinct components of MW progenitor analogues and our own Galaxy. In all of these efforts, we emphasize the need for increased transparency around the choice of Solar scale and abundance methodology, in order to maximize the scientific return of studies attempting to couple extragalactic and Galactic chemical abundances. 

\section*{Acknowledgments}
The authors warmly thank the referee for their helpful comments and suggestions that greatly improved this paper. The ChemZz collaboration is incredibly grateful to the organisers and support staff of the Lorentz Center workshop ``Gravitational waves: a new ear on the chemistry of galaxies'', where the idea for this paper was born. SM thanks Magda Arnaboldi for her very helpful feedback and Vasily Belokurov for discussions which helped progress this project. TMS and FC acknowledge support from a UKRI Frontier Research Guarantee Grant (PI Cullen; grant reference: EP/X021025/1). JLS acknowledges support from the Royal Society (URF\textbackslash R1\textbackslash191555). CK acknowledges funding from the UK Science and Technology Facility Council through grant ST/Y001443/1. TS gratefully acknowledges the support of the NSF-Simons AI-Institute for the Sky (SkAI) via grants NSF AST-2421845 and Simons Foundation MPS-AI-00010513 and of NASA grants 22-ROMAN22-0055 and 22-ROMAN22-0013. This work relies heavily on the \texttt{Astropy} \citep{astropy1, astropy2}, \texttt{SciPy} \citep{scipy}, \texttt{NumPy} \citep{numpy} and \texttt{Matplotlib} \citep{matplotlib} libraries and \texttt{Jupyter} notebooks \citep{jupyter}. 

This work makes use of data from the GALAH survey, observed using the Anglo-Australian Telescope on the unceded territory of the Gamilaraay people. We acknowledge the traditional custodians of the land and pay our respects to elders past and present. 

Some of the data presented herein were obtained at Keck Observatory, which is a private 501(c)3 non-profit organization operated as a scientific partnership among the California Institute of Technology, the University of California, and the National Aeronautics and Space Administration. The Observatory was made possible by the generous financial support of the W. M. Keck Foundation. The authors also wish to recognise and acknowledge the very significant cultural role and reverence that the summit of Maunakea has always had within the Native Hawaiian community. We are most fortunate to have the opportunity to conduct observations from this mountain. 

\section*{Data Availability}
Data for the Cosmic Noon sample is included in Table \ref{tab:cosmic_noon_abund}. Abundances for the Milky Way sample, Local Group and M~31 are all already publicly available. 

\bibliography{lorentz-chemzz}
\bibliographystyle{aasjournal}

\label{lastpage}

\end{document}